\pdfoutput=1
\documentclass[bibyear,traditabstract]{aa}
\usepackage{graphicx}
\usepackage{txfonts}
\usepackage{natbib}
\bibpunct{(}{)}{;}{a}{}{,} 
\begin{document}

\title{Full non-LTE spectral line formation \\ III. The case of 
  a two-level atom with broadened upper level}

\author{ M. Sampoorna\inst{1} \and F. Paletou\inst{2,3} \and V. Bommier\inst{4}
\and T. Lagache\inst{2,3}}

          \institute{Indian Institute of Astrophysics, Koramangala, Bengaluru
  		560034, India
             \email{sampoorna@iiap.res.in}
            \and
	  Universit\'e de Toulouse, UPS-Observatoire
            Midi-Pyr\'en\'ees, Cnrs, Cnes, Irap, Toulouse, France
            \and
          Cnrs,  Institut de Recherche en Astrophysique et
            Plan\'etologie, 14 av. E. Belin, F--31400 Toulouse, France
            \and
	    LESIA, Observatoire de Paris, Universit\'e PSL, Sorbonne 
	    Universit\'e, Universit\'e Paris Cit\'e, CNRS, Meudon, France
            }

   \date{Received xxx; accepted yyy}


   \abstract{In the present paper we consider the full nonlocal 
   thermodynamic equilibrium (non-LTE) radiation transfer problem. 
   This formalism allows us to account for deviation from equilibrium 
   distribution of both the radiation field and the massive particles. 
   In the present study two-level atoms with broadened upper level represent 
   the massive particles. In the absence of velocity-changing 
   collisions, we demonstrate the analytic equivalence of the full non-LTE 
   source function with the corresponding standard non-LTE partial frequency 
   redistribution (PFR) model. We present an iterative method based 
   on operator splitting techniques to numerically solve the problem at hand. 
   We benchmark it against the standard non-LTE transfer problem for a 
   two-level atom with PFR. We illustrate the deviation of the 
   velocity distribution function of excited atoms from the equilibrium 
   distribution. We also discuss the dependence of the emission profile and 
   the velocity distribution function on elastic collisions and 
   velocity-changing collisions. }


   \titlerunning{Full non--LTE spectral line formation III.}

   \maketitle

\section{Introduction}
\label{sec-intro}
In a series of two papers \citep{pp21,psp23}, we have been revisiting
the problem of so-called ``full'' nonlocal thermodynamic
equilibrium (non-LTE) radiation transfer originally formulated by
\citet{ox86}. This formalism not only accounts for deviation of the
radiation field from the Planckian equilibrium distribution, but also
for the deviation of the velocity distribution of massive particles
from the Maxwellian equilibrium distribution. While \citet{pp21} focused 
on setting the stage, \citet{psp23} considered the numerical solution of 
the problem for the case of coherent scattering in the atom's frame, which 
corresponds to scattering on a two-level atom with (putative) infinitely 
sharp energy levels. In the present paper, we further extend these works 
for scattering on a two-level atom with infinitely sharp lower level and 
a more realistic broadened upper level (which may already be 
suitable for the modeling of strong resonance lines). 

The full non-LTE radiation transfer formalism is based on the 
kinetic theory of particles and photons \citep{ox86}. 
In particular, it is founded on a semi-classical description 
of light scattering in spectral lines. As described in \citet{hm14}, the 
semi-classical picture combines concepts from classical and the more exact 
quantum mechanical description of the problem at hand, thereby providing a 
very intuitive and compelling approach to the problem. Clearly, the 
semi-classical picture is not a self-consistent theory and therefore contains 
a number of not fully defined concepts. However, it has been very successful in 
describing several of the line scattering mechanisms in astrophysical 
conditions (see the above-cited books for details).
For the problem considered in this paper, the poorly defined concepts from the 
physical point of view are those related to the rate of velocity-changing 
collisions and a clear distinction between the elastic and velocity-changing 
collisions. However, despite this, we introduce separate rates for the elastic 
and velocity-changing collisions. Although it is not completely clear how they 
would be evaluated for actual cases, and even a proper quantum-mechanical 
definition of these quantities is uncertain, their introduction and usage in 
the present paper is fully in the spirit of the semi-classical picture that
we adopt here. Indeed this semi-classical theory provides a way to treat the 
problem, namely including a self-consistent determination of the velocity 
distribution of atoms in the upper level of the transition. For a more 
detailed outline of the semi-classical picture, we refer the reader to 
\citet[][see their Chapter 10, specifically pp. 291--294]{hm14}.

In the full non-LTE formalism, the kinetic 
equation for the velocity distribution of the massive particles 
(namely, the atoms or ions and free electrons) and that for the photons 
(namely, the radiative transfer equation for the intensity of the radiation 
field) have to be formulated and solved simultaneously and self-consistently. 
Since the velocity distribution functions (VDFs) of the atomic levels are not 
known a priori, the absorption and emission profiles that enter the radiative 
transfer equation need to be obtained by convolving the corresponding atomic 
quantities with the VDFs, wherein the velocity of the massive particle 
is measured in the observer's frame. An evaluation of the need to use this 
formalism had remained unexplored because of the numerical complexity involved 
in its implementation. The present series of papers aim to clarify this 
question through detailed numerical calculations. For this purpose, we have 
embarked upon developing suitable numerical techniques to implement this 
formalism. As a first step \citet{psp23} considered the two-distribution 
problem, namely the intensity of the radiation field and the VDF of the 
excited atoms are the only two distributions that need to be 
determined simultaneously and self-consistently. In other words, two-level 
atoms represent the massive particles with their lower (ground) level having 
the equilibrium Maxwellian distribution \citep{ox86,pp21}. 
Furthermore, stimulated emission was neglected, and 
the free electrons that are responsible for inelastic collisions between the 
two levels of the atom were also assumed to obey the equilibrium 
Maxwellian distribution. In the present paper we continue to consider this 
two-distribution problem, with the important difference that the lower level 
of the atom  continues to be infinitely sharp, while the upper level is 
broadened. In this case, the atomic absorption profile is a Lorentzian and 
the atomic emission profile already depends on the radiation field. This 
introduces some difficulties in the numerical solution of the corresponding 
full non-LTE problem, namely we need to accurately compute Voigt-like function 
which involves a Lorentzian function in its integrand \citep{pp20}. Following a 
method developed previously by \citet{vb97a,vb97b}, in the present paper we 
device a simple and efficient technique to compute such an integral involving 
Lorentzian function.

The basic equations of the two-distribution problem are detailed 
in \citet[][see also \citealt{pp21}]{psp23}. Hence, we do not repeat them 
here, and only the equations relevant to the two-level atom with broadened 
upper level are discussed. Outline of the present paper is as follows. In
Section~\ref{sec-profiles} we discuss the explicit forms of the
absorption and emission profiles for a two-level atom with broadened
upper level. The full non-LTE source function is presented in
Section~\ref{sec-source}, wherein we also demonstrate the analytic
equivalence with the corresponding standard non-LTE source function in
the absence of velocity-changing collisions.
In Section~\ref{sec-coll-clarify}, we describe and 
clarify the three different types of collisions (namely, the inelastic, 
elastic, and velocity-changing collisions) considered in this paper. 
Section~\ref{sec-num-meth} is devoted to the numerical method of
solution for the full non-LTE problem considered here. Numerical
results are illustrated and discussed in
Section~\ref{sec-num-res}. Conclusions are presented in
Section~\ref{sec-conclu}.

\section{The absorption and emission profiles}
\label{sec-profiles}
Like in the standard non-LTE formalism 
\citep[see e.g.,][who also adopt the semi-classical picture]{hm14}, the 
absorption and emission profiles (and also the frequency redistribution 
functions) are first determined in the atomic rest frame, and then transformed 
to the observer's frame to account for the Doppler motion of the atoms in a 
stellar atmosphere. In the standard non-LTE formalism with complete 
frequency redistribution (CFR), the VDF of all the atomic levels is assumed 
to be the equilibrium Maxwellian distribution, while when partial frequency 
redistribution (PFR) is included this assumption is limited to only the 
lower level. However, the standard non-LTE formalism with PFR does not 
provide access to the VDF of the upper level. This is provided by the full 
non-LTE formalism. Therefore, in this section 
we discuss the absorption and emission profiles first in the atomic frame 
and then in the observer's frame.

For the case of a two-level atom with broadened upper level, the atomic
absorption profile is given by 
\citep[see Appendix~B.2 of][see also \citealt{pp21}]{ox86} 
\begin{equation}
\alpha_{12}(\xi) = {\frac{\delta_w}{\pi}} {\frac{1}{(\xi-\nu_0)^2+\delta_w^2}},
\label{aabs12}
\end{equation}
where $\xi$ is the photon frequency in the atomic frame, $\nu_0$ is
the line-center frequency, and the damping width
$\delta_w=(A_{21}+Q_{I}+Q_E)/(4\pi)$, with $A_{21}$ being the
Einstein coefficient for spontaneous emission or radiative deexcitation rate, 
$Q_{I}$ the inelastic collisional deexcitation rate 
\citep[denoted as $C_{21}$ in][]{pp21}, and 
$Q_E$ the total elastic collision rate. 

The absorption profile in the observer's frame is given by 
\begin{equation}
\varphi_{\nu} = \int_{\vec{u}}
\alpha_{12}(\nu-\Delta\nu_D\,{\vec{u}}\cdot{\vec {\Omega}}) f_1({\vec{u}})d^3{\vec{u}}, 
\label{ofabs12}
\end{equation}
where ${\vec{u}}$ is the atomic velocity\footnote{Here the velocity of
the atom is measured in the observer's frame.} normalized to the thermal
velocity ($\varv_{\rm th}=\sqrt{2kT/M}$, with $k$ being the Boltzmann constant, 
$T$ the temperature, and $M$ the mass of the atom), ${\vec{\Omega}}$ is the 
propagation direction of the ray,
and $\Delta\nu_D$ is the Doppler width. In this paper, we do not 
account for bulk velocities resulting from mass motion of the massive 
particles. In other words only Doppler motion of atoms is taken into account, 
so that the corresponding velocities are in the nonrelativistic regime, 
wherein only the photon frequency is subject to Lorentz transformation 
between the atomic rest frame and the observer's frame, while the photon 
direction remains unchanged \citep[i.e., aberration is neglected; see also 
Eqs.~(2.4.3a)--(2.4.3e) in page 54 of][]{ox86}. Therefore, in the above 
equation we have used the Fizeau–Doppler relationship 
\citep[see Eq.~(9) of][]{pp21},
which relates the frequencies in the atomic ($\xi$) and observer's
($\nu$) frames.  Furthermore, $f_1$ represents the 
VDF of the lower level of the atom. In the
weak radiation field regime $f_1$ can be assumed to be the equilibrium
distribution, namely a Maxwellian $f^M$ \citep{ox86,pp21}. With this
assumption it is straight forward to show that the resulting
absorption profile in the observer's frame is a normalized Voigt
function $\varphi(x)=H(a,x)$, where $a=\delta_w/\Delta\nu_D$ and
$x=(\nu-\nu_0)/\Delta\nu_D$, with $\nu_0$ being the line-center
frequency.

In the following subsections we discuss the emission profile first in the 
atomic frame (see Section~\ref{sec-emiprof-af}) and then in the observer's 
frame (see Section~\ref{sec-emiprof-of}). 

\subsection{The atomic emission profile}
\label{sec-emiprof-af}
The explicit form of the atomic emission profile in the absence of 
velocity-changing collisions is given in Eq.~(B.2.26) of \citet{ox86} 
and in Eq.~(4.32) of \citet{hos83a}. Although, the notations used in the 
above-said references are somewhat different, it can be readily shown that 
both the mentioned expressions are identical. In the presence of 
velocity-changing collisions, the atomic emission profile is given in 
\citet[][see their Section~4.1]{hos83b}. In their notations, this atomic 
emission profile is given by 
\begin{equation}
\eta_{21}(\xi,\tau) = {\frac{B_{12}I^\ast_{12}j_{121}(\xi,\tau)+
[S_{\!12}+\gamma_2(n_2/n_1)]r_{12}(\xi)}
{B_{12}I^\ast_{12}+S_{\!12}+\gamma_2(n_2/n_1)}},
\label{eta21}
\end{equation}
where $B_{12}$ is the Einstein coefficient for radiative absorption, $S_{\!12}$
is the collisional excitation rate, $\gamma_2$ is the velocity-changing 
collision rate, $n_1$ and $n_2$ are the number density of the atoms 
in lower and upper levels, respectively, and $\tau$ is the line 
center optical depth. In Eq.~(\ref{eta21}), the generalized redistribution 
function $r_{12}(\xi)=\alpha_{12}(\xi)$ \citep[see Eq.~(6.3.55) of][]{ox86}, 
the quantity $I^\ast_{12}$ is given by 
\begin{equation}
I^\ast_{12} = \int r_{12}(\xi)\, I(\nu,{\vec{\Omega}},\tau) \,	d\xi\,,
\label{i12}
\end{equation}
and 
\begin{equation}
j_{121}(\xi,\tau) = {\frac{1}{I^\ast_{12}}}\int r_{121}(\xi',\xi)\, 
I(\nu',{\vec{\Omega}}',\tau) \,d\xi'\,,
\label{j121hos}
\end{equation}
with $\int d\xi = \oint\int d\nu\, d{\vec{\Omega}}/(4\pi)$. In the above 
equations, $I(\nu, \vec{\Omega}, \tau)$ is the specific intensity, and 
$r_{121}(\xi^\prime, \xi)$ is the generalized atomic redistribution 
function \citep{hos83a}, which describes the joint probability of absorbing 
a photon of frequency $\xi^\prime$ and spontaneously re-emitting a photon of 
frequency $\xi$. 

We remark that the quantity $I^\ast_{12}$ introduced by \citet{hos83a,hos83b} 
is the same as $I_{12}$ introduced in Eq.~(B.2.20) of \citet{ox86}. In 
the notations of \citet[][see also \citealt{psp23}]{pp21}, we readily identify 
$I^\ast_{12}=J_{12}({\vec{u}},\tau)$, $S_{\!12}=C_{12}$, and 
$\gamma_2=Q_V$\footnote{In \citet{pp21} and \citet{psp23}, the 
velocity-changing collision rate $Q_V$ was denoted by $Q_2$. This changed  
notation has been adopted to avoid the possible confusion with the notation 
$Q_2$ used in \citet[][see e.g., their Eq.~(10.151) in page 327]{hm14} for the 
elastic collision rate.}. Therefore, we re-write Eq.~(\ref{eta21}) 
in the present notations as follows
\begin{equation}
\eta_{21}(\xi,\tau) = {\frac{B_{12}J_{12}({\vec{u}},\tau)j_{121}(\xi,\tau)+
[C_{12}+Q_V(n_2/n_1)]\alpha_{12}(\xi)}{B_{12}J_{12}({\vec{u}},\tau)+C_{12}
+Q_V(n_2/n_1)}}, 
\label{eta21q2}
\end{equation}
where $J_{12}({\vec{u}},\tau)$ is defined as 
\citep[see e.g., Eq.~(8) of][]{pp21}
\begin{equation}
J_{12}(\vec{u},\tau) = \oint{ { {d\Omega} \over {4 \pi} } } \int_0^{\infty}{
\alpha_{12}(\nu - \Delta\nu_D\,\vec{u} \cdot \vec{\Omega}) 
I(\nu, \vec{\Omega}, \tau)\, d\nu} \, .
\label{j12}
\end{equation}
Also, in our notations, the quantity $j_{121}(\xi,\tau)$ takes the
form
\begin{equation}
j_{121}(\xi,\tau)={\frac{1}{J_{12}(\vec{u},\tau)}} 
\oint{ { {d\Omega^\prime} \over {4 \pi} } } \int_0^{\infty}{
r_{121}(\xi^\prime, \xi) 
I(\nu^\prime, \vec{\Omega}^\prime, \tau)\, d\nu^\prime} \, .
\label{j121}
\end{equation}
The atomic frequencies appearing in the above equation are related to their 
observer's frame counterparts through Fizeau–Doppler relationship, namely, 
$\xi^\prime=\nu^\prime - \Delta\nu_D\,\vec{u} \cdot \vec{\Omega}^\prime$ and 
$\xi=\nu - \Delta\nu_D\,\vec{u} \cdot \vec{\Omega}$. 

\subsection{The observer's frame emission profile}
\label{sec-emiprof-of}
The emission profile in the observer's frame is given by
\begin{equation}
\psi_{\nu}({\vec \Omega},\tau) = \int_{\vec{u}}
\eta_{21}(\nu - \Delta\nu_D\,\vec{u} \cdot \vec{\Omega},\tau) 
f_{2}({\vec{u}},\tau) d^3{\vec{u}}. 
\label{psi}
\end{equation}
The VDF of the upper-level including the velocity-changing 
collisions is given in Eq.~(3) of \citet{psp23}. However, for the purpose 
of deriving the emission profile in the observer's frame, we use the 
VDF as given in Eq.~(19) of \citet{pp21}, namely
\begin{equation}
f_{2}({\vec{u}},\tau) = {\frac{n_1}{n_2}}{\frac{B_{12}J_{12}({\vec{u}},\tau)+
C_{12}+Q_V(n_2/n_1)}{A_{21}+C_{21}+Q_V}} f^M({\vec{u}}).
\label{f2col}
\end{equation}
Substituting Eqs.~(\ref{eta21q2}) and (\ref{f2col}) into Eq.~(\ref{psi}), the 
latter takes the form
\begin{eqnarray}
\psi_{\nu}({\vec{\Omega}},\tau)&=& \int_{\vec{u}} \Bigg\{ {\frac{n_1}{n_2}}
\left[{\frac{B_{12}J_{12}({\vec{u}},\tau)j_{121}(\xi,\tau)+
C_{12}\alpha_{12}(\xi)}{A_{21}+C_{21}+Q_V}} \right] 
\nonumber \\&&
+{\frac{Q_V\alpha_{12}(\xi)}{A_{21}+C_{21}+Q_V}} \Bigg\}
f^{M}({\vec{u}}) d^3{\vec{u}}.
\label{psicoll}
\end{eqnarray}
Following Section~5 of \citet{pp21}, we assume LTE values for $n_1$
(namely, $n_1=n^\ast_1$), and introduce the normalized populations
$\bar n_2=n_2/n^\ast_2$. Substituting for $j_{121}$ (cf.
    Eq.~(\ref{j121})), the first term in the flower brackets of the
above equation can be re-written as
\begin{equation*}
{\frac{1}{\bar n_2}} {\frac{1}{1+\zeta}} \left[
\varepsilon \alpha_{12}(\xi) +(1-\varepsilon)
\oint{ { {d\Omega^\prime} \over {4 \pi} } } \int_0^{\infty}{
r_{121}(\xi^\prime, \xi) 
I(\nu^\prime, \vec{\Omega}^\prime, \tau)\, d\nu^\prime}
\right], 
\end{equation*}
where we have made use of Eqs.~(23)--(27) of \citet{pp21}, and
$I(\nu^\prime, \vec{\Omega}^\prime, \tau)$ is now normalized to the
Planck function in the Wien limit. In the above expression 
\begin{equation}
\varepsilon = {\frac{Q_I}{A_{21}+Q_I}},
\label{eps}
\end{equation}
is the usual collisional destruction 
probability and the quantity 
\begin{equation}
\zeta={\frac{Q_V}{A_{21}+Q_I}}, 
\label{zeta}
\end{equation}
gives the amount of velocity-changing collisions. Similarly the 
last term in the flower brackets of Eq.~(\ref{psicoll}) reduces to $\zeta
\alpha_{12}(\xi) / (1+\zeta)$. Using Eq.~(29) of \citet{pp21}, the 
emission profile can be re-written as
\begin{eqnarray}
\psi_{\nu}({\vec{\Omega}},\tau)&\!\!\!\!=&\!\!\!\!\!\!\!\! \int_{\vec{u}} 
f^{M}({\vec{u}}) d^3{\vec{u}}\Bigg\{ 
{\frac{\zeta}{1+\zeta}}\alpha_{12}(\xi) +{\frac{1}{1+\zeta}} 
{\frac{1}{\varepsilon+(1-\varepsilon){\mathcal J}_{12}(\tau)}} 
\nonumber \\&&\!\!\!\!\!\!\!\!\!\!\!\!\!\!\!\!\!\!\!\!\!\!\!\!\!\!\!\!\!\times
\left[
{\varepsilon \alpha_{12}(\xi)+(1-\varepsilon)
\oint{ { {d\Omega^\prime} \over {4 \pi} } } \!\!\int_0^{\infty}{\!\!
r_{121}(\xi^\prime, \xi) 
I(\nu^\prime, \vec{\Omega}^\prime, \tau)\, d\nu^\prime}}
\right] \Bigg\},
\label{psicoll1}
\end{eqnarray}
where 
\begin{equation}
{\cal{J}}_{12}(\tau) = \int_{\vec{u}}{ \bar{J}_{12}(\vec{u}, \tau)
f^{M}(\vec{u}) d^3 \vec{u}} \,.
\label{j12cal}
\end{equation}
In the above equation $\bar{J}_{12} (\vec{u},\tau) = {J}_{12}
(\vec{u},\tau)/ {\mathcal B}_W$, with ${\mathcal B}_W$ denoting the Planck
function in the Wien limit. Substituting for ${J}_{12}(\vec{u},\tau)$ from 
Eq.~(\ref{j12}) and using Eq.~(\ref{ofabs12}), it can be easily shown that 
\begin{equation}
{\cal{J}}_{12}(\tau) = \oint{ { {d\Omega} \over {4 \pi} } }
\int_0^{\infty}{\varphi_{\nu}\,
I(\nu, \vec{\Omega}, \tau)\, d\nu}\,,
\label{j12caln}
\end{equation}
is simply the frequency integrated mean intensity (also referred to as 
``CFR scattering integral'') appearing in the standard non-LTE
problem. Furthermore, we may readily identify that $\int
r_{121}(\xi',\xi)f^{M}({\vec{u}}) d^3{\vec{u}} =
R_{121}(\nu^\prime,{\vec{\Omega}}^\prime,\nu,{\vec{\Omega}})$, namely the 
angle-dependent generalized redistribution function in the 
observer's frame. Furthermore, using Eq.~(\ref{ofabs12}), we obtain the 
emission profile in the observer's frame as
\begin{eqnarray}
\psi_{\nu}({\vec{\Omega}},\tau)&\!\!\!\!=&\!\!\!\!\!
{\frac{\zeta}{1+\zeta}} \varphi_{\nu}+
{\frac{1}{1+\zeta}}
{\frac{1}{\varepsilon+(1-\varepsilon){\mathcal J}_{12}(\tau)}} 
\nonumber \\&&\!\!\!\!\!\!\!\!\!\!\!\!\!\!\!\!\!\!\!\!\!\!\!\!\!\!\!\!\!\!\!\times
\left[\varepsilon \varphi_{\nu}+(1-\varepsilon) 
\oint{ { {d\Omega^\prime} \over {4 \pi} } } \!\!\int_0^{\infty}{\!\!\!\!
R_{121}(\nu^\prime,{\vec{\Omega}}^\prime,\nu,{\vec{\Omega}})
I(\nu^\prime, \vec{\Omega}^\prime, \tau)\, d\nu^\prime}\right]. 
\label{psicoll2}
\end{eqnarray}
We remark that the emission profile given above is the same as 
that originally derived in \citet[][see their Eq. (4.15)]{hos83b}, although 
we have presented it in a slightly different form and also using the 
notations adopted in this paper. Furthermore, \citet{hc86} have demonstrated 
that the emission profile derived from the semi-classical picture of 
\citet{hos83b} fully agrees with that derived from a quantum mechanical 
approach of \citet{cbbh82}.

In the present paper, we consider the angle-averaged emission 
profile, namely 
\begin{equation}
\psi(x,\tau) = 
\oint{ { {d\Omega} \over {4 \pi} } } \psi(x,{\vec{\Omega}},\tau),
\label{psix}
\end{equation}
wherein we have transformed from real frequency $\nu$ to the non-dimensional 
frequency $x$. The resulting angle-averaged emission profile is given by 
\begin{eqnarray}
\psi(x,\tau)&\!\!\!\!=&\!\!\!\!\!
{\frac{\zeta}{1+\zeta}} \varphi(x)+
{\frac{1}{1+\zeta}}
{\frac{1}{\varepsilon+(1-\varepsilon){\mathcal J}_{12}(\tau)}} 
\nonumber \\&&\!\!\!\!\!\!\!\!\!\!\!\!\!\!\!\!\!\!\!\!\!\!\!\!\!\!\times
\left[\varepsilon \varphi(x)+(1-\varepsilon) 
\oint{ { {d\Omega^\prime} \over {4 \pi} } } \!\!\int_{-\infty}^{+\infty}{\!\!
R_{121}(x^\prime,x)
I(x^\prime, \vec{\Omega}^\prime, \tau)\, dx^\prime}\right]. 
\label{psicollx}
\end{eqnarray}
For a two-level atom with broadened upper level and in the absence 
of velocity-changing collisions, the generalized redistribution function 
$R_{121}$ is given by the observer's frame counterpart of the usual atomic 
frame PFR function derived by \citet[][see \citealt{mm73} and \citealt{mih78} 
for the corresponding observer's frame expression]{osc72}. In the presence of 
velocity-changing collisions, the explicit form of the PFR function $R_{121}$ 
has been derived in \citet[][see their Eqs.~(3.15) and (3.16)]{hc86} starting 
from the quantum mechanical approach of \citet{cbbh82}. In this paper, we adopt 
this PFR function, which in our notations takes the form\,:
\begin{equation}
R_{121}(x^\prime,x)
= \gamma_{\rm coh,\,V}\ 
R_{\rm II-A}(x^\prime,x) + 
(1-\gamma_{\rm coh,\,V})\ 
R_{\rm III-A}(x^\prime,x),
\label{r121}
\end{equation}
where $R_{\rm II-A}$ and $R_{\rm III-A}$ are respectively the type-II and 
type-III angle-averaged PFR functions of \citet{hum62}, and the 
coherence fraction is given by
\begin{equation}
\gamma_{\rm coh,\,V} = {\frac{A_{21}+Q_{I}+Q_V}{A_{21}+Q_{I}+Q_{E}}}.
\label{gamcohv}
\end{equation}
Clearly, in the absence of velocity-changing collisions, the coherence fraction 
as well as the PFR function become identical to those derived by \citet{osc72}.

To clearly bring out the departure of the emission profile from CFR,
we may re-write Eq.~(\ref{psicollx}) as
\begin{eqnarray}
\psi(x,\tau)&=&
\varphi(x)+ {\frac{1}{1+\zeta}} 
{\frac{(1-\varepsilon)}{\varepsilon+(1-\varepsilon){\mathcal J}_{12}(\tau)}} 
\nonumber \\&&\!\!\!\!\!\!\!\!\!\!\!\!\!\!\!\!\!\!\!\!\!\!\!\!\!\!\times
\left[
\oint{ { {d\Omega^\prime} \over {4 \pi} } } \!\!\int_{-\infty}^{+\infty}{\!\!
R_{121}(x^\prime,x)
I(x^\prime, \vec{\Omega}^\prime, \tau)\, dx^\prime} - {\mathcal J}_{12}(\tau)
\varphi(x)\right]. 
\label{psicollxn}
\end{eqnarray}
The above equation can easily be deduced from Eq.~(\ref{psicollx}) by simply 
adding and subtracting unity to $\zeta$ appearing in the numerator of the 
first term of that equation.\footnote{We remark that 
Eq.~(\ref{psicollxn}) can more easily be related to Eq.~(4.15) of 
\citet{hos83b}.}

\section{The source function}
\label{sec-source}
In the full non-LTE formalism, the source function for a two-level atom with 
broadened upper level is of the form \citep[see Eq.~(7) of][]{psp23}
\begin{equation}
  S(x, \tau) = \left[ \varepsilon + (1 - \varepsilon)
    {\cal{J}}_{12}(\tau) \right] \left[ {{\psi(x, \tau)} \over
      {\varphi(x)}} \right] \, ,
	\label{sxt}
\end{equation}
where $\psi(x, \tau)$ is given by Eq.~(\ref{psicollx}) or 
Eq.~(\ref{psicollxn}) and $\varphi(x)$ is the normalized Voigt function. 

In the absence of velocity-changing collisions (namely, $\zeta=0$), 
the source function~(\ref{sxt}) reduces to the corresponding expression for 
the standard non-LTE PFR model for a two-level atom with broadened 
upper level. This can be easily realized, by noting that the emission profile 
for $\zeta=0$ (cf. Eq.~(\ref{psicollx})) takes the following form
\begin{eqnarray}
\psi(x,\tau)&\!\!\!\!=&\!\!\!\!\!
{\frac{1}{\varepsilon+(1-\varepsilon){\mathcal J}_{12}(\tau)}} 
\nonumber \\&&\!\!\!\!\!\!\!\!\!\!\!\!\!\!\!\!\!\!\!\!\!\!\!\!\!\!\times
\left[\varepsilon \varphi(x)+(1-\varepsilon) 
\oint{ { {d\Omega^\prime} \over {4 \pi} } } \!\!\int_{-\infty}^{+\infty}{\!\!
R_{121}(x^\prime,x)
I(x^\prime, \vec{\Omega}^\prime, \tau)\, dx^\prime}\right]. 
\label{psicollx-z0}
\end{eqnarray}
Substituting Eq.~(\ref{psicollx-z0}) into Eq.~(\ref{sxt}), we readily obtain 
\begin{equation}
 S(x, \tau) = 
\varepsilon +(1-\varepsilon) 
\oint{ { {d\Omega^\prime} \over {4 \pi} } } \int_{-\infty}^{+\infty}{
\left[{\frac{R_{121}(x^\prime,x)}{\varphi(x)}}\right]
I(x^\prime, \vec{\Omega}^\prime, \tau)\, dx^\prime},
\label{sxt-z0}
\end{equation}
thereby establishing the equivalence of the source function derived from the 
full and standard non-LTE models. This is an important result that 
justifies the use of numerically relatively simpler standard non-LTE formalism 
in the absence of velocity-changing collisions.

In the presence of velocity-changing collisions, the source
function is obtained by substituting Eq.~(\ref{psicollxn}) into
Eq.~(\ref{sxt}), namely
\begin{eqnarray}
S(x, \tau)& =& \varepsilon + (1 - \varepsilon){\cal{J}}_{12}(\tau) + 
{\frac{(1-\varepsilon)}{1+\zeta}} 
\nonumber \\&&\!\!\!\!\!\!\!\!\!\!\!\!\!\!\!\!\!\!\!\!\!\!\!\!\!\!\times
\left\{
\oint{ { {d\Omega^\prime} \over {4 \pi} } } \int_{-\infty}^{+\infty}{
\left[{\frac{R_{121}(x^\prime,x)}{\varphi(x)}}\right]
I(x^\prime, \vec{\Omega}^\prime, \tau)\, dx^\prime} - {\mathcal J}_{12}(\tau)
\right\}.
\label{sxt-z}
\end{eqnarray}
We recall that as in \citet{pp21} and \citet{psp23} the source
function and intensity are normalized to the Planck function in the
Wien limit. When velocity-changing collisions are significant
(namely, $\zeta \gg 1$), the third term in Eq.~(\ref{sxt-z}) is
negligible and the source function tends to the CFR source function.

\section{A clarification on collisions}
\label{sec-coll-clarify}
In the present paper, we consider three types of collisions\,:
\begin{itemize}
\item inelastic collisions,
\item elastic collisions, and
\item velocity-changing collisions.
\end{itemize}
For the massive particles (two-level atoms in the present paper), we 
distinguish between\,:
\begin{itemize}
\item the ``internal variables'' (the levels energy and quantum 
numbers), and 
\item the ``external variables'' (the position and velocity).
\end{itemize}

The inelastic collisions are those responsible for collisional 
transitions between the lower and upper levels. They enter the kinetic (or 
statistical equilibrium) equation for atoms as inducing transitions, 
which further lead to the $\varepsilon$ factor in the radiative transfer 
equation. In an inelastic collision, the internal variables change, whereas
the external variables may or may not change. 

The elastic collisions do not modify the level populations. They 
are responsible for line broadening. In a strictly elastic collision both the 
internal and external variables remain unchanged before and after the 
collision. In a weakly elastic collision, the internal variables, namely the 
energy value of the level and the quantum numbers may change (for example 
collision between fine-structure or hyperfine structure levels, or between 
Zeeman sublevels in the presence of a weak magnetic field), while the external 
variables remain unchanged. It is also important to note that elastic 
collisions between degenerate Zeeman sublevels also lead to change in 
internal variables (as it changes the magnetic quantum number before and 
after collisions, although not the atomic energy). Indeed in the present 
paper we consider isolated spectral line and do not consider magnetic fields. 
Thus the Zeeman sublevels are degenerate. If an elastic collision changes the 
Zeeman sublevel (namely, the magnetic quantum number $M$ is changed to $M'$, 
with $M \ne M'$), then it will have a polarizing or, more frequently, 
depolarizing effect (note that polarization results from unequal populations 
between the Zeeman sublevels). On the contrary, if an elastic collision does 
not change the Zeeman sublevel (namely, $M \to M$), then it only broadens the 
line, namely, it is only a line broadening collision. As clarified in 
\citet{sb19} line broadening and depolarizing collisions are different and 
both contribute to the line broadening and hence to the elastic collision 
rate $Q_E$.

Apart from contributing to line broadening, the elastic collisions 
take part in the frequency redistribution of the scattered radiation in the 
line as shown for e.g., in \citet[][see also \citealt{vb97a,vb97b,sb19}]{osc72}.
In particular, they are responsible for CFR in the atomic frame. 
In the full non-LTE formalism, both these effects of elastic collisions are 
included following the works of 
\citet[][see also Appendix B.2 of \citealt{ox86}]{osc72} 
and \citet{hc86} in the absence and presence of velocity-changing collisions, 
respectively. This is done by including $Q_E$ in the total damping width of 
the absorption profile (see Eq.~(\ref{aabs12})) and using the appropriate 
PFR function of \citet{osc72} and \citet{hc86} for the generalized 
redistribution function (see Eq.~(\ref{r121})).

On the other hand, in a collision if the internal variables remain 
unchanged, while the external variables such as atomic velocity (in particular 
its modulus) is changed, then such type of collisions are called 
velocity-changing collisions. We think that since the internal variable is 
unchanged, \citet{ox86} refers to velocity-changing collision as elastic. 
However, since the atomic velocity is changed during the collision, we feel 
that it may not be appropriate to refer to this type of collision as 
``elastic''. Indeed \citet{ll04} do not refer to them as elastic. Also, most 
likely, the velocity-changing collisions represented by $Q_V$ includes both 
elastic and inelastic contributions. Furthermore, this type of collisions are 
close collisions or strong collisions with small impact parameter. For the 
two-distribution problem considered in this paper, the velocity-changing 
collisions enter the kinetic equation for excited atoms as a relaxation 
term as they are responsible for relaxing the VDF of the upper level to 
its equilibrium distribution function. This relaxation term is 
of the form given in Eq.~(7) of \citet[][see also Eq.~(6.3.12) in page 167 
of \citealt{ox86}]{pp21}. 

In general, a given collision can modify both the internal and external 
variables. Those collisions which modify only the internal variables 
are most likely long-range collisions (obviously, short-range collisions can 
also modify the internal variables), and those collisions which modify the 
external variable (namely, velocity) are strong or (only) short-range 
collisions. In this respect, $Q_V$ is a part of $Q_E$ 
\citep[see Sections~4.1.1 and 4.1.2 of][see also Section~II(a) of \citealt{hc86}]{vb16a}.
Even if velocity-changing collisions are part of the line-broadening 
collisions, the method of calculation of $Q_V$ is not the same as the method 
of calculation of $Q_E$, because $Q_E$ addresses the atomic internal variables, 
while $Q_V$ addresses the atomic external variables, although the 
colliding particles are the same. The velocity-changing collisions are 
atom-atom collisions, while the elastic and inelastic collisions may also be 
caused by electron-atom collisions (in addition to atom-atom collisions). 

\citet{hc86} show that when lower state interaction is negligibly 
small (namely, the collisional scattering amplitude of the lower level is much 
smaller than that of the upper level), the total elastic collision rate $Q_E$
associated with the upper level can be decomposed into two parts, one 
corresponding to only phase changes without change in velocity (denoted 
$q_E$) and other corresponding to both phase and velocity changes 
\citep[denoted as $\nu$ by][]{hc86}. Such a decomposition is well suited 
for resonance lines to which our present full non-LTE approach is applicable. 
Therefore, following \citet{hc86}, we write $Q_E=q_E+Q_V$ after identifying 
their $\nu$ as our $Q_V$. Furthermore, \citet{hc86} show that $q_E=\alpha Q_E$ 
and $Q_V=(1-\alpha)Q_E$ with $\alpha$ in the range $0$ to $1$. They also give 
an estimate of $(1-\alpha)$. When $m\ll M$ (with $m$ denoting the mass of the 
perturber and $M$ denoting the mass of the radiator), they show that 
$(1-\alpha) = (m/M)^2$, and when $m \sim M$, they estimate $(1-\alpha)=0.1$. 

It is known that the phase-changing elastic collision rate $q_E$ can 
be obtained from Van der Waals approximation or using the more precise 
semi-classical theory developed in 1990s by Anstee, Barklem, and O'Mara 
\citep[the so-called ABO theory; see e.g.,][and references cited therein]{bo98,bao98}. 
As for the velocity-changing collision rate $Q_V$, a precise calculation of the 
corresponding collision cross-section will depend on the atomic species under 
consideration. \citet{ll04} give only a rough order of magnitude for this 
cross-section (to be on the order of 10 to 100 $\pi\,a_0^2$, with $a_0$ being 
the Bohr radius). Specific calculations would be necessary for a better 
precision. There has been a laboratory study by \citet{bvb78}, who 
measure the effects of the velocity-changing collisions between the 
excited Kr atoms and the He and/or Ar perturbers. Here the authors 
show that a ``hard-sphere'' collision model is suitable for interpreting 
their experimental measurements. According to this model the collisional 
cross-section is given by $\pi(r_A+r_B)^2$, where $r_A$ and $r_B$ are the 
radii of the atoms participating in the collision. The atomic radii for any 
atomic specie (including also the ions) are listed by 
\citet[][see page 45]{allen73}, which the authors used for 
computing the collisional cross-section for Kr*-He and Kr*-Ar 
collisions. In the present paper, all the three collisional rates are 
assumed to be input parameters, and hence we do not compute them using the 
collisional dynamics. Consequently, we do not determine the velocity 
distribution of the colliders.

In order to evaluate the importance of velocity-changing collisions 
in a stellar atmosphere, \citet{ll04} provide a way to estimate the critical 
density of the perturbers or colliders (see their Eq. (13.6) in page 694). 
This is done by comparing an order of magnitude rate for velocity-changing 
collisions (given by $nqv$, where $n$ is the number density of the perturbers, 
$q$ is the  cross section for velocity-changing collisions, and $v$ is the 
average velocity of perturbers relative to the atom) with the rate for 
spontaneous emission (namely, $A_{21}$). The density for which these 
two rates are nearly the same gives the critical density\,: 
$n_c \approx 7.8 \times 10^{16} A_{21} / q / \sqrt{T}$ (in units of cm$^{-3}$).
Here $A_{21}$ is in units of $10^7$ s$^{-1}$, temperature $T$ is in units of 
$10^4$ K, and $q$ is in units of $\pi\,a_0^2$. 
For densities larger than this critical density velocity-changing 
collisions are significant. \citet{ll04} estimate $q$ to be in a 
range\footnote{We have verified that  
the range of $q$ values suggested by \citet{ll04} more or less agrees with 
those determined from the hard-sphere collision model. For example, using the 
atomic radii listed in \citet[][see page 45]{allen73}, we find $q$ for H-H 
collision to be 7, He-H collision to be 12.9, and Cs-Cs collision to be 137 
(note that Cs has the largest atomic radius).}
that is rarely larger than $10$ to $100$.
Using this, \citet{vb16a} provided an estimate of the critical density of 
colliders to be on the order of $10^{20}$ cm$^{-3}$. We redid this estimate 
for the same set of parameters as used by her (namely, $A_{21} =1$), except 
for the temperature (chosen here to be $T=0.5$), which is not mentioned in 
her Section 4.1.1. We find the critical density $n_c$ to be in the range 
$1.103\,\times\,10^{16}$ to $1.103\,\times\,10^{15}$ cm$^{-3}$. This is about 
4 to 5 orders of magnitude smaller than the one mentioned in \citet{vb16a}. 
Since the collisions with neutral hydrogen is expected to be the dominant 
source of both elastic and velocity-changing collisions in a stellar 
atmosphere, we compare $n_c$ with the hydrogen density in the solar 
photosphere, which  is on the order of $10^{17}$ cm$^{-3}$. Clearly, the 
velocity-changing collisions are important in the lower solar atmosphere, and 
hence have to be accounted for. Furthermore, the collisional destruction 
probability $\varepsilon$ (which depends on the inelastic collision rate 
$C_{21}$ or $Q_I$; see Eq.~(\ref{eps})) also becomes non-negligible in the 
lower solar atmosphere \citep[see e.g., Fig. 2(c) of][]{lsaetal10}. Finally, 
as discussed in Section 4.1.1. of \citet{vb16a}, the elastic collision rate 
$Q_E$ is also significant in the lower solar atmosphere. Thus the present full 
non-LTE formalism is applicable in the lower solar atmosphere when 
velocity-changing collisions are important. In the upper solar atmosphere 
wherein velocity-changing collisions are negligible, one may use the 
numerically relatively simpler standard non-LTE PFR formalism.

\section{The numerical method of solution} 
\label{sec-num-meth}
We solve the full non-LTE transfer problem for the case of a two-level
atom with broadened upper level using a modified version of the
Accelerated Lambda Iteration (ALI) method developed by \citet{pa95}
for the corresponding standard PFR model. Here we present this ALI
method in some detail, focusing on the changes brought about by the
full non-LTE nature of the problem at hand.  As in the standard PFR 
model, the source function given by Eq.~(\ref{sxt-z}) is iterated
until convergence using the Approximate Lambda Operator (ALO) which
is chosen to be the diagonal of the full lambda operator
\citep{oab86}.

In an one-dimensional planar atmosphere considered here, the radiation field 
is axisymmetric. Thus the specific intensity depends only on the inclination 
$\theta_r$ of the ray about the atmospheric normal. In other words 
$I(x, \vec{\Omega}, \tau)=I(x, \mu, \tau)$, where $\mu=\cos\theta_r$. Thus, 
the formal solution of the radiative transfer equation can be stated as 
\begin{equation}
I_{x\mu}=\Lambda_{x\mu}[S_x],
\label{fs-lambda}
\end{equation}
where for notational convenience, we have suppressed the dependence on optical 
depth, and the dependence on frequency and angular variables appear as 
subscript. Moreover $\Lambda_{x\mu}$ denotes the frequency and angle-dependent 
integral operator. Given an estimate of the source function at the $n$th 
iteration, the iterative scheme will be given by 
\begin{equation}
S^{(n+1)}_x=S^{(n)}_x+\delta S^{(n)}_x,
\label{snp1}
\end{equation}
where $\delta S^{(n)}_x$ is the iterative correction on the source
function.  Using the operator splitting technique \citep{can73},
namely, $\Lambda_{x\mu}=
\Lambda^{\ast}_{x\mu}+(\Lambda_{x\mu}-\Lambda^{\ast}_{x\mu})$ with
$\Lambda^{\ast}_{x\mu}$ being the ALO chosen here to be diagonal of
the full lambda operator after \citet{oab86}, and following a rather
standard procedure \citep[][see also \citealt{sj10}]{pa95}, we arrive
at the following expression for the iterative correction
\begin{eqnarray}
&&\!\!\!\!\!\!\!\!\!\!
\delta S^{(n)}_x - (1-\varepsilon)\int_0^{\infty} \varphi_x\,
\Lambda^{\ast}_{x}[\delta S^{(n)}_x]\,dx - {\frac{(1-\varepsilon)}{(1+\zeta)}} 
\nonumber \\ &&\!\!\!\!\!\!\!\!\!\!\times
\left\{\int_0^{\infty}\left[{\frac{R_{121}(x^\prime,x)}{\varphi_x}}\right]
\Lambda^{\ast}_{x^\prime}[\delta S^{(n)}_{x^\prime}]\,dx^\prime - 
\int_0^{\infty} \varphi_x\,\Lambda^{\ast}_{x}[\delta S^{(n)}_x]\,dx\right\}
\nonumber \\ &&\!\!\!\!\!\!\!\!\!\!
 = r^{(n)}_x\,.
\label{deltasn}
\end{eqnarray}
For deducing the above equation, we have used Eq.~(\ref{j12caln}) and
also the fact that for a static atmosphere the radiation field is
symmetric about the line-center, so that only half profile can be
considered. In the above equation, the frequency dependent ALO is
given by
\begin{equation}
\Lambda^{\ast}_{x}=\int_{-1}^{+1}{\frac{d\mu}{2}}\,\Lambda^{\ast}_{x\mu},
\label{lamx}
\end{equation}
and the residual $r^{(n)}_x$ has the form
\begin{eqnarray}
r^{(n)}_x & =& \varepsilon + (1 - \varepsilon){\cal{J}}^{(n)}_{12}(\tau) +
{\frac{(1-\varepsilon)}{1+\zeta}} 
\nonumber \\&&\!\!\!\!\!\!\!\!\!\!\!\!\!\!\!\!\!\!\!\times
\left\{
\int_{0}^{\infty}{
\left[{\frac{R_{121}(x^\prime,x)}{\varphi_x}}\right]\Lambda_{x^\prime}[S^{(n)}_{x^\prime}]
\, dx^\prime} 
- {\mathcal J}^{(n)}_{12}(\tau)\right\} - S^{(n)}_x\,, 
\label{residual}
\end{eqnarray}
where ${\cal{J}}^{(n)}_{12}(\tau)$ and the integral involving the 
angle-averaged PFR function are obtained from the formal solver 
using the $n$th iterate source function. For this purpose, we use the 
short-characteristic method of \citet[][see also \citealt{lambetal16}]{ok87}. 
At each iteration the system of linear equations~(\ref{deltasn}) can be 
resolved using either a frequency-by-frequency (FBF) method or a core-wing 
method \citep{pa95}. In the following subsections we briefly describe both 
these methods for the full non-LTE case considered here. 

\subsection{Frequency-by-Frequency method}
\label{sec-fbf}
For a given depth point the system of equations~(\ref{deltasn}) consists of 
$N_x$ number of linear equations, with $N_x$ representing the number of 
frequency points. In matrix form this system of linear equations can be 
written as 
\begin{equation}
{\bf A}\,\delta{\vec{S}}^{(n)} = {\vec{r}}^{(n)}\,,
\label{fbf}
\end{equation}
where at each depth point, $\delta{\vec{S}}^{(n)}$ and ${\vec{r}}^{(n)}$ are 
vectors of length $N_x$ and ${\bf A}$ is a matrix of dimension 
$N_x\times N_x$. Following \citet{pa95} we solve Eq.~(\ref{fbf}) using 
the LU decomposition scheme \citep[see e.g.,][]{pressetal86}. As the FBF method 
involves matrix manipulations such as inversion and multiplication, it is 
computationally somewhat expensive when compared to the core-wing method 
presented in the following subsection.

\subsection{Core-wing method}
\label{sec-core-wing}
Based on the behavior of the type-II PFR function of \citet{hum62}, 
a core-wing method was proposed by \citet{pa95} that allowed the computation of 
system of linear equations~(\ref{deltasn}) through simple algebraic 
manipulations, thereby considerably reducing the computational costs involved. 
In this method, the type-II PFR function is approximated by 
CFR in the line core and coherent scattering in the wings for the 
computation of the source function corrections. An extension of this method 
for the type-III PFR function was given by \citet{flurietal03}, 
wherein this function is approximated by CFR in the line core and 
set to zero in the wings. We apply 
both the above mentioned core-wing approximations to the $R_{121}$ function 
appearing in Eq.~(\ref{deltasn}). Furthermore, in Eq.~(\ref{deltasn}) we make 
the approximation of computing the frequency integral involving the absorption 
profile $\varphi_x$ only in the line core and set it to zero in the wings. 
This approximation is similar to the core-wing approximation made for 
type-III PFR function. With these approximations, we can easily 
deduce the following core-wing approximation for Eq.~(\ref{deltasn})\,:  
\begin{equation}
\delta S^{(n)}_x = {\frac{r^{(n)}_x+(1-\alpha_x)\,\Delta T^{\rm core}}
{1-(1-\varepsilon)\,\alpha_x\,\Lambda^\ast_x/(1+\zeta)}}\,,
\label{core-wing}
\end{equation}
where $\alpha_x$ is the core-wing separation coefficient given by
\begin{equation}
\label{sepa-coeff}
\alpha_x = 
\begin{cases}
0 & \text{in the core } (x\leqslant 3.5), \\ 
\gamma_{\rm coh,\,V}\,{R_{\rm II-A}(x,x)}/{\varphi_x} & \text{in the wings } 
(x > 3.5),
\end{cases}
\end{equation}
and 
\begin{equation}
\Delta T^{\rm core} = (1-\varepsilon)\,\int_{\rm core} \varphi_x\,
\Lambda^\ast_x[\delta S^{(n)}_x]\,dx\,, 
\label{dtcore}
\end{equation}
which can be easily evaluated as described in 
\citet[][see their Section~5.1]{pa95}. As in \citet{sj10}, we find that 
when the elastic collision rate $Q_E$ and/or velocity-changing 
collision rate $Q_V$, are large the 
approximation of setting the type-III PFR function and the 
${\mathcal J}_{12}(\tau)$ integral to zero in the wings leads to convergence 
problems. In such cases we use the CFR approximation throughout the 
line profile. This typically occurs for $Q_E/A_{21}>1$ and/or 
$Q_V/A_{21}>1$, when 
the medium is optically thick or semi-infinite. We have verified that both 
the core-wing and FBF methods give identical results. Therefore, all 
the solutions presented in this paper are computed with the core-wing 
method. 

\subsection{Numerical computation of $\bar{J}_{12}({\vec{u}},\tau)$}
\label{sec-jbaru}
The full non-LTE formalism gives us access to the VDF $f_2$ of the upper level, 
which depends on $\bar{J}_{12}({\vec{u}},\tau)$ 
\citep[see Eq.~(3) of][]{psp23}. In the case of a two-level atom with 
broadened upper level, the computation of $\bar{J}_{12}({\vec{u}},\tau)$ is 
numerically more complex than the coherent scattering case considered in 
\citet{psp23}. This is because, in the case of coherent scattering the 
integrand in $\bar{J}_{12}({\vec{u}},\tau)$ contained a delta function 
(see their Eq.~(1)), while the integrand in the present case involves a 
Lorentzian function (see Eq.~(\ref{j12})). Integrals involving Lorentzian 
are known to be notoriously difficult to evaluate due to the sharp peaked 
nature of the Lorentzian function. Therefore, we need to devise 
a suitable method to evaluate such integrals accurately. 
Here we describe such a numerical method following \citet{vb97a,vb97b}. 

In terms of the adimensional frequency $x$, the 
quantity $\bar{J}_{12}({\vec{u}},\tau)$ is given by (cf. Eq.~(\ref{j12}))
\begin{equation}
\bar{J}_{12}(\vec{u},\tau) = \oint{ { {d\Omega} \over {4 \pi} } } 
\int_{-\infty}^{+\infty}{
{\frac{a}{\pi}} {\frac{1}{(x-\vec{u} \cdot \vec{\Omega})^2+a^2}}\,
I(x, \mu, \tau)\, dx}. 
\label{j12-x}
\end{equation}
The dot product of the velocity vector ${\vec{u}}$ with the ray direction 
${\vec{\Omega}}$ is evaluated using Eq.~(9) of \citet{psp23}, which requires 
us to construct the corresponding quadratures for polar angles $\theta_u$ and 
$\theta_r$. Since the radiation field is axisymmetric, it is sufficient to 
construct the quadrature directly for the azimuth difference $(\phi_r-\phi_u)$. 
The dot product ${\vec{u}} \cdot {\vec{\Omega}}$ can take both positive 
and negative values. Thus, the frequency integral in Eq.~(\ref{j12-x}) has to 
include the entire range from $-\infty$ to $+\infty$. The intensity 
for the negative $x$ values can be easily obtained from the corresponding 
positive values using the symmetry relation. We have used thirteen 
Gauss-Legendre nodes for the direction cosines corresponding to both the ray 
and velocity vector in the $[0,1]$ domain, and a twenty-point equally spaced 
quadrature for their azimuth difference $(\phi_r-\phi_u)$ in the 
$[0,2\pi]$ domain. For each pair of $\theta_r$, $\theta_u$, and 
$(\phi_r-\phi_u)$, we first perform the frequency integral and then the 
angular integration. 

Evaluating integrals involving Lorentzian function poses accuracy
issues \citep[see
  e.g., Fig.~1 of][]{pp20}.  In the present paper, we apply a method
originally developed by \citet{vb97a,vb97b} to compute the
angle-dependent type-III PFR function of \citet{hum62}, which is known
to involve an integration over the Lorentzian function \citep[see
  e.g., Eq.~(61) of][]{vb97b}. Following her method, we compute the
frequency integral in Eq.\,(\ref{j12-x}) by the trapezoidal method with
varying integration steps. The integration begins from the center of
the Lorentzian (namely, at $x=\vec{u} \cdot \vec{\Omega}$), and
proceeds symmetrically thereafter. The integration step is originally
set as one tenth of the damping width $a$ and multiplied by 1.05 at
each step of the integration (namely, a geometric progression). As
discussed above, the quantity $\vec{u} \cdot \vec{\Omega}$ is
evaluated as described in \citet[][see their Eq.~(9)]{psp23}. Thus the
integration needs to be performed for each value of $u$ and $\gamma$
(which is the cosine of the angle between the velocity vector and the
ray direction). Since the frequency integration step size is varied as
described above, the intensity computed on a standard frequency grid
used for radiative transfer needs to be interpolated at every step of
the frequency integration. We use the spline interpolation for this
purpose.

The method described above to compute $\bar{J}_{12}({\vec{u}},\tau)$
is however, somewhat slow. For a typical case of semi-infinite
atmosphere with 10 points per decade, 65 frequency points, and the
angular quadrature mentioned above it requires about 67 minutes of
computing time on a Intel(R) Xeon(R) Gold 5122 processor with 3.6GHz,
10.4 GT/s clock speed.  Clearly computing this quantity and
subsequently the VDF $f_2$ of the upper level at every iteration would
be computationally very expensive. However, since the ALI method
described above does not require us to compute
$\bar{J}_{12}({\vec{u}},\tau)$ and $f_2$ at every iteration (see
Eq.~(\ref{deltasn}), we compute these quantities once the ALI solution
has converged, which typically takes 15 seconds of computing
time. 

\section{The numerical results} 
\label{sec-num-res}
In this section we first validate our iterative method by reproducing
the benchmark result of \citet{hum69}, and then illustrate the new
quantities, namely the VDF of the upper level and the emission
profiles together with source function with and without elastic 
and/ or velocity-changing collisions. We also illustrate a comparison 
of the normally emergent intensity profiles for the cases of a two-level 
atom with infinitely sharp upper and lower levels considered in \citet{psp23}, 
and a two-level atom with broadened upper level considered in this paper,
along with the corresponding CFR standard non-LTE models. For the 
numerical studies presented here, we consider a one-dimensional, isothermal, 
semi-infinite, planar atmosphere of total optical thickness at line center 
$T=10^6$ and $\epsilon=10^{-4}$. The radiative width of the upper level
parameterized as $a_R=A_{21}/(4\pi\Delta\nu_D)$ is chosen to be
$10^{-3}$.  It is related to the total damping parameter via
$a=a_R[1+(Q_{I}+Q_E)/A_{21}]$. Unless otherwise mentioned, 
both the rates $q_E/A_{21}$ and $Q_V/A_{21}$ are set to zero.

\subsection{Validation}
\label{sec-validate}
\begin{figure}[]
  \includegraphics[width=9cm, angle=0]{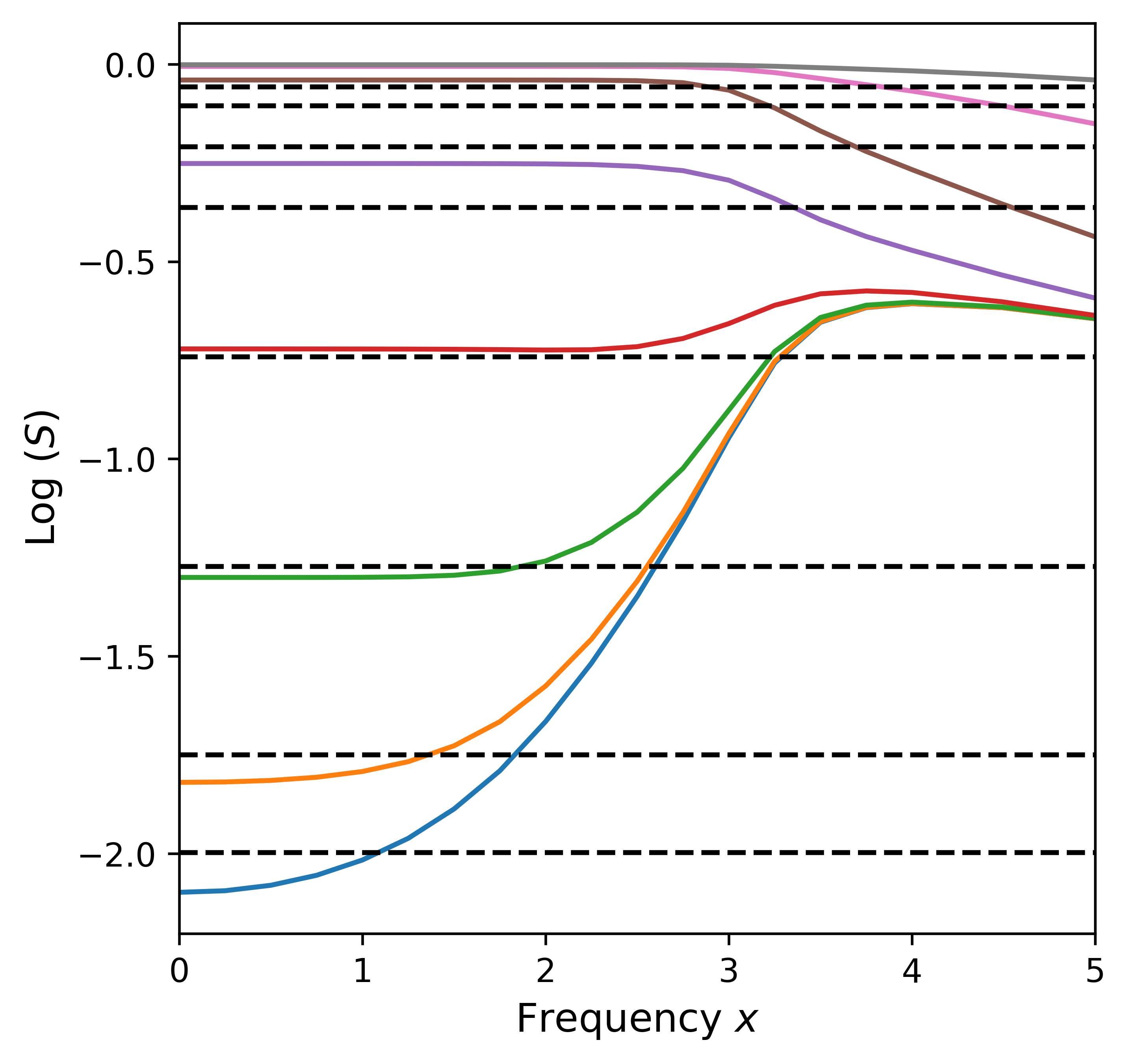}
	\caption{Validation of our iterative method for full non-LTE
          transfer problem \citep[compare with Fig.~3c
            of][]{hum69}. The normalized source function is displayed
          as a function of frequency at different line center optical
          depths within the atmosphere, namely, at $\tau= 0, \,1,
          \,10, \,100,\,10^3,\,10^4$. For comparison, we also show the
	  corresponding CFR source function (constant with frequency)
          as dashed lines. 
	  }
  \label{fig1-hum69-validate}
\end{figure}

For the atmospheric model described above, the standard non-LTE source
function for the $R_{II-A}$ PFR model is illustrated in Fig.~3c of
\citet{hum69}. For the purpose of validating our numerical method we
reproduce this benchmark result in our
Figure~\ref{fig1-hum69-validate}, which displays the source function
$S(x,\tau)$ for different optical depths as a function of frequency.
These solutions are computed with the ALI method presented in
Section~\ref{sec-num-meth} together with the core-wing method (cf.
Section~\ref{sec-core-wing}) for calculating the source function
corrections. A comparison of our Fig.~\ref{fig1-hum69-validate} with
Fig.~3c of \citet{hum69}, clearly shows that our numerical method
satisfactorily reproduces the benchmark solutions thereby validating
our iterative method. This is expected, as the source function derived
from full non-LTE formalism is equivalent to the corresponding
standard non-LTE PFR model when velocity-changing collisions 
are neglected (cf. Section~\ref{sec-source}).

\begin{figure}[]
  \includegraphics[width=9.0 cm, angle=0]{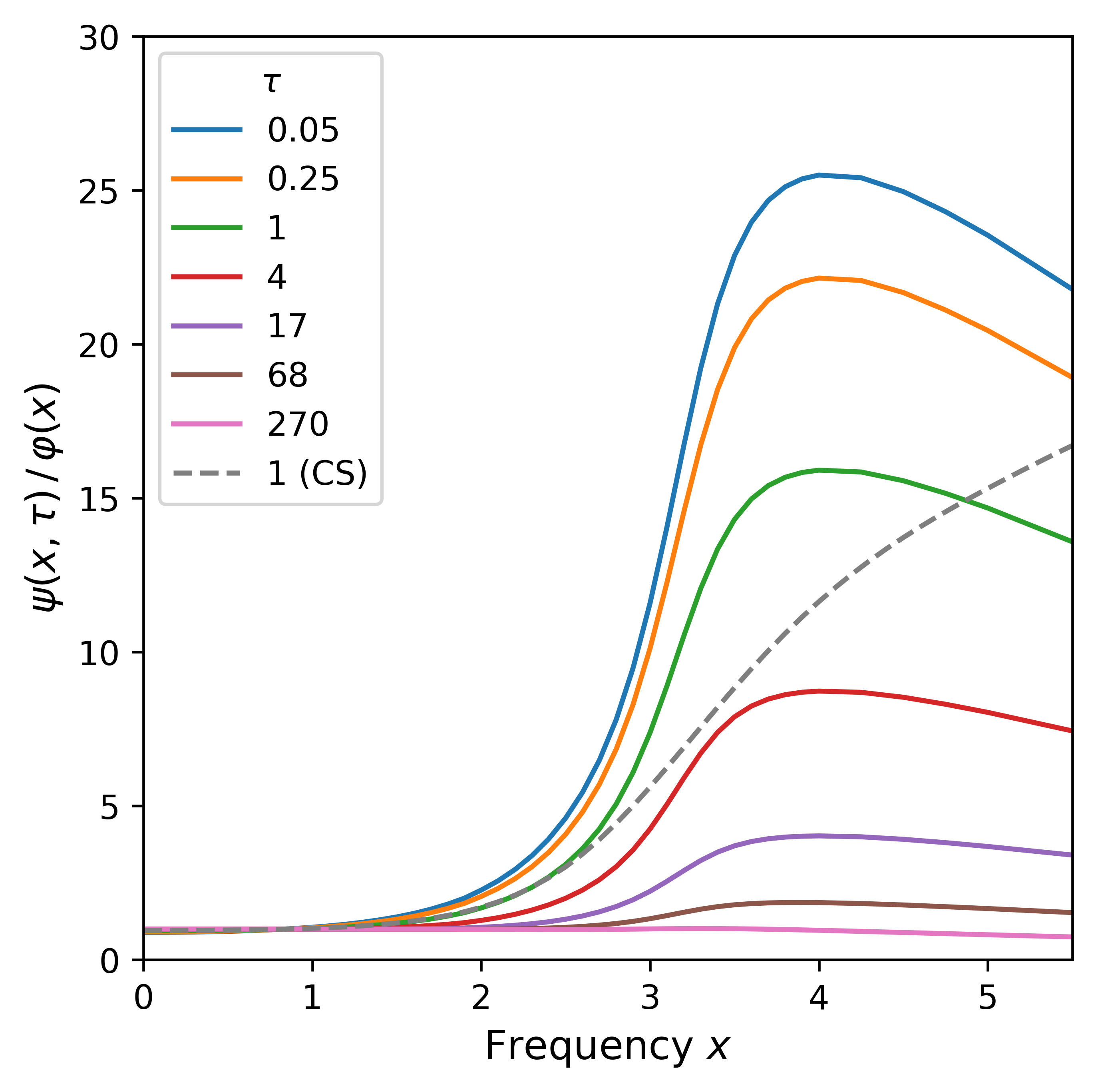}
	\caption{Departure of emission profile $\psi(x, \tau)$ from
	  CFR for the case of scattering on a two-level atom with
          radiatively broadened upper level. Different lines
          correspond to $\psi(x, \tau)/\varphi(x)$ at different
          line-center optical depths within the atmosphere (indicated
          in the figure legend). For comparison
          $\psi(x,\tau=1)/\varphi(x)$ corresponding to scattering on a
          two-level atom with infinitely sharp upper and lower levels
          (namely, coherent scattering -- CS in the atomic frame) is
          shown as dashed line.}
  \label{fig2-psi-hum69}
\end{figure}

\begin{figure}[]
  \includegraphics[width=9.0 cm, angle=0]{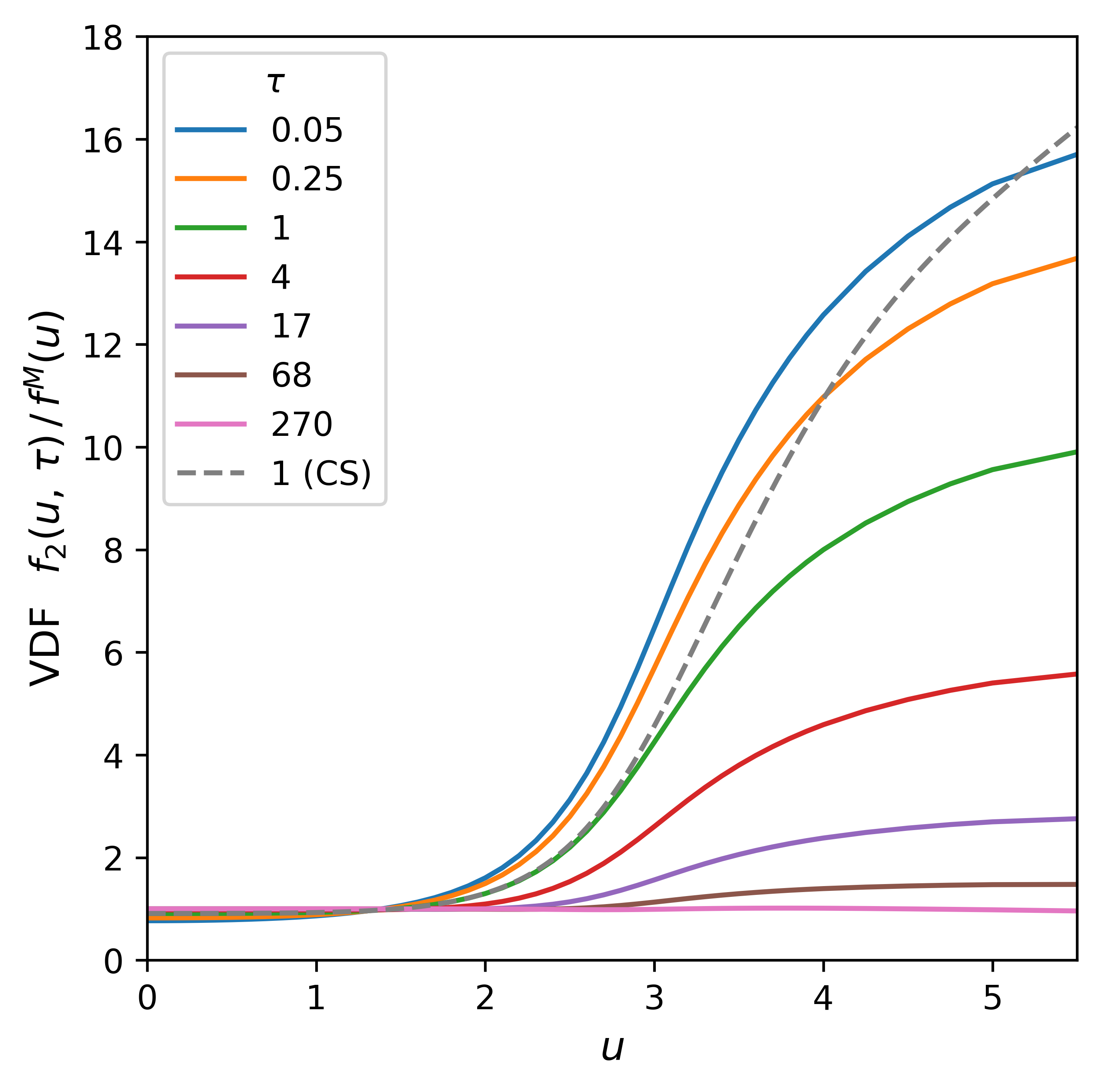}
	\caption{Departure of the VDF of the naturally broadened upper level 
	($f_2(u,\tau)$) of a two-level atom from the Maxwellian equilibrium 
	distribution $f^M (u)$ at different line center optical depths within 
	the atmosphere (indicated in the figure legend). For comparison the 
	corresponding quantity at $\tau=1$ for the CS case is shown as dashed 
	line. Clearly, overpopulation of $f_2$ at large $u$'s is relatively 
	more in the CS case than in the present case of two-level atom with 
	naturally broadened upper level.}
  \label{fig3-f2-hum69}
\end{figure}

Unlike the standard non-LTE PFR model considered by \citet{hum69}, 
the full non-LTE model considered here gives access to the 
VDF of the upper level. The emission profile on the other hand can be 
obtained from both the above-mentioned formalisms. However, it is rarely 
shown in the literature. Therefore, in this paper we illustrate both the 
emission profile and the VDF of the upper level. Figures\,\ref{fig2-psi-hum69}
and \ref{fig3-f2-hum69} exhibit respectively the ratios
$\psi(x,\tau)/\varphi(x)$ and $f_2(u,\tau)/f^M(u)$ for different line
center optical depths within the
atmosphere. Figure\,\ref{fig3-f2-hum69} in the present paper is
equivalent to Fig.~3\ in \citet{psp23}, but for the case of scattering
on a two-level atom with naturally broadened upper level. For the ease
of comparison, in Figs.~\ref{fig2-psi-hum69} and \ref{fig3-f2-hum69},
we also show as dashed lines the corresponding quantities at $\tau=1$
for coherent scattering (CS) in atomic frame (namely, the case of
two-level atom with infinitely sharp lower and upper levels)
considered in \citet{psp23}.

For the standard non-LTE CFR model, the emission and the absorption
profiles are identical \citep{hm14}. Thus to bring out the departure
of the emission profile from CFR, we plot in Fig.~\ref{fig2-psi-hum69}
the ratio $\psi(x,\tau)/\varphi(x)$ at different line center optical
depths within the atmosphere. Clearly, the emission profile departs
from CFR for $x>1$. As the optical depth increases this departure 
from CFR decreases. Furthermore, the differences in
$\psi(x,\tau)/\varphi(x)$ between the present and CS cases are
significant (compare green solid and gray dashed lines in
Fig.~\ref{fig2-psi-hum69}).

Because here we consider angle-averaged emission profile, the VDF of
the excited atom depending only on the modulus of velocity $u$ is
illustrated. Like in the CS case, deviation from the Maxwellian
distribution is significant for $u>2$ and for optical depths close to
the surface, which then decreases with increasing optical depth
\citep[compare the Fig.~\ref{fig3-f2-hum69} here with Fig.~3
  of][]{psp23}.  However, unlike the CS case the overpopulation of
excited level for $u>2$ is relatively smaller in the present case of
two-level atom with naturally broadened upper level (compare green
solid and gray dashed lines in Fig.~\ref{fig3-f2-hum69}). 
We remark here that departure of the VDF of the upper level from 
the Maxwellian distribution was also obtained by 
\citet[][see her Section~5.3]{vb16b} through 
a self-consistent solution of the statistical equilibrium equations for 
each velocity class of the velocity-dependent atomic density matrix elements 
and the radiative transfer equation for the polarized radiation in the case 
of Na\,{\sc i} D$_1$ and D$_2$ lines. This departure can be attributed to 
the radiative processes between the interacting atom and the incident radiation 
field which is spectrally structured (namely, non-flat) within the radiative 
width of the upper level.

\begin{figure}[]
  \includegraphics[width=9.0 cm, angle=0]{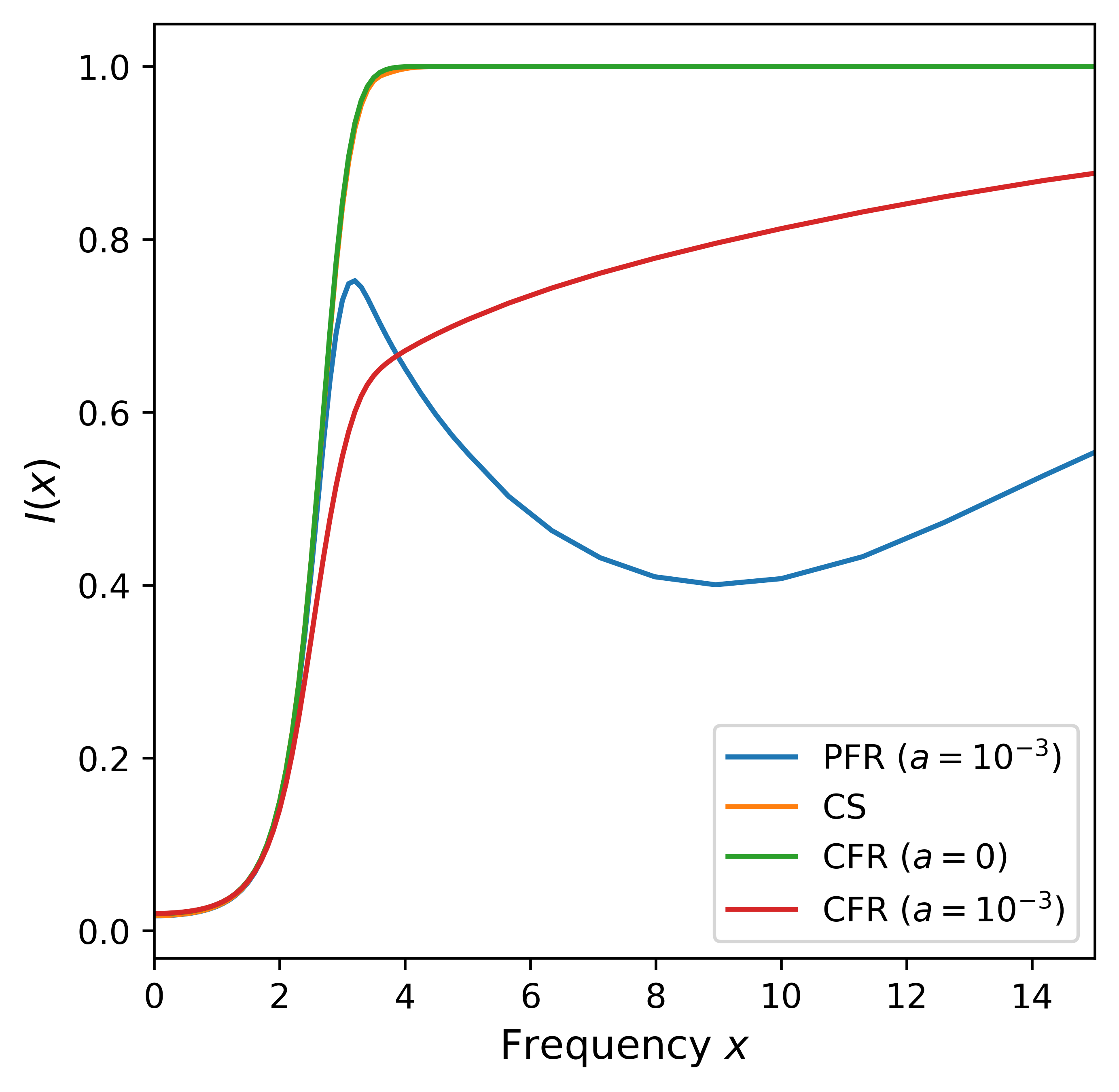}
	\caption{Comparison of normally emergent intensity computed
          using full non-LTE model for a two-level atom with
          infinitely sharp levels (CS) and radiatively broadened upper
	  level (PFR) and standard non-LTE model with CFR. 
	  Notice that the intensity for CS and the
	  corresponding CFR ($a=0$) case nearly coincide.}
  \label{fig4-int-hum69}
\end{figure}

 Figure\,\ref{fig4-int-hum69} displays a comparison of the
  normally emergent intensity for the CS and the present case of a
  two-level atom with radiatively broadened upper level. We also
plot the corresponding CFR cases, namely for the damping parameter
$a=0$ and $a=10^{-3}$. While the CS and the corresponding CFR ($a=0$)
cases nearly coincide (compare orange and green lines in
Fig.~\ref{fig4-int-hum69}), significant differences are seen in the
wings for $x>3$ between the CS and PFR cases (compare green and blue
lines). Moreover, the PFR intensity differs significantly from the
corresponding CFR intensity for $x>2$ (compare blue and red lines in
Fig.~\ref{fig4-int-hum69}). In particular, we recover the lowering / dip 
of emergent intensity in the wings before finally reaching the continuum 
level, a well-known effect of PFR ($R_{\rm II-A}$).

\subsection{Impact of velocity-changing collisions ($Q_V/A_{21}$)}
\label{sec-vcoll}

\begin{figure}[]
  \includegraphics[width=9cm, angle=0]{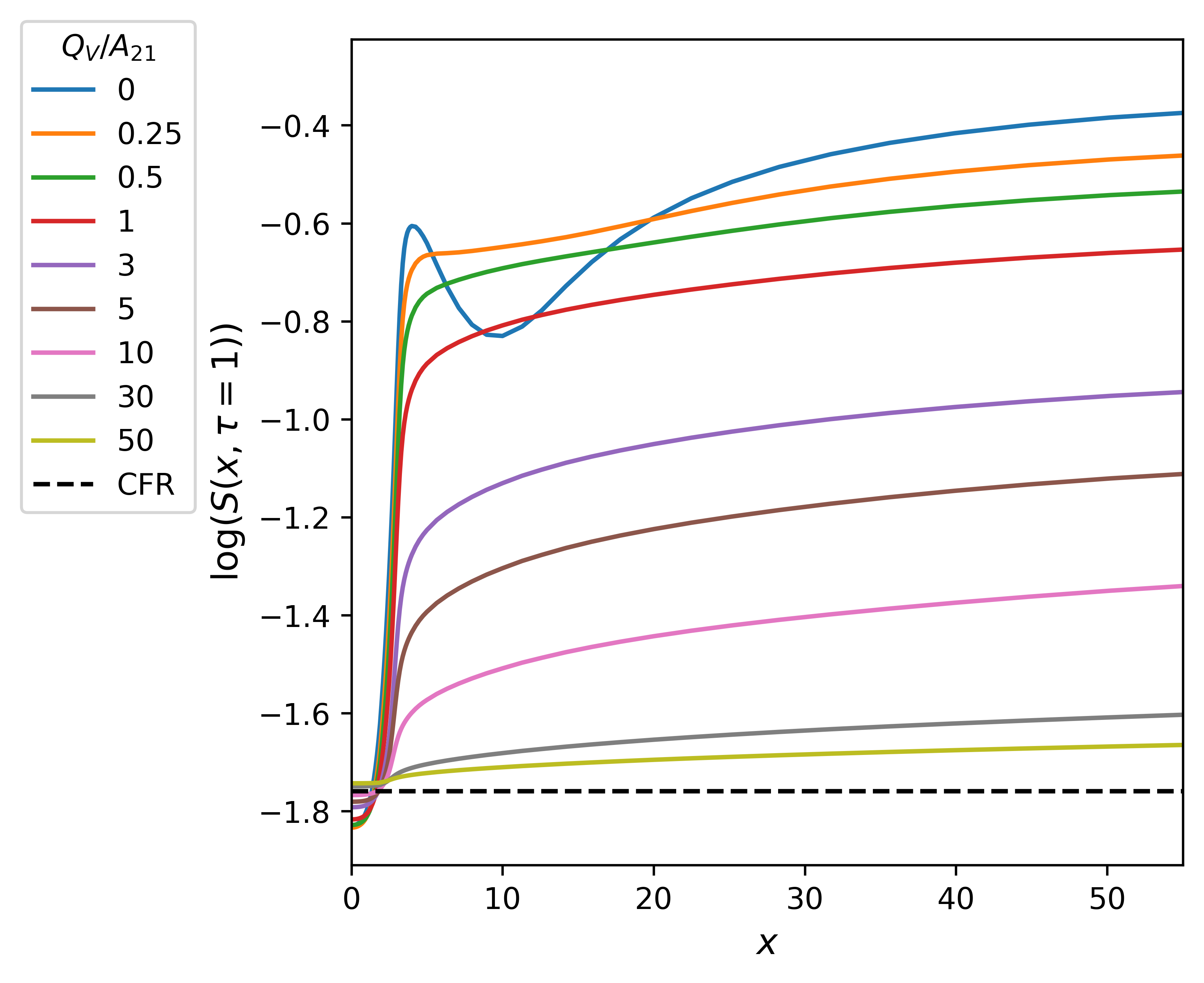}
	\caption{Influence of velocity-changing collisions 
	on the normalized source function at $\tau=1$. As 
	expected, with increasing values of $Q_V/A_{21}$, the 
	source function approaches the CFR limit, shown as 
	horizontal dashed line. 
    }
  \label{fig5-s-zeta}
\end{figure}

\begin{figure}[]
  \includegraphics[width=9cm, angle=0]{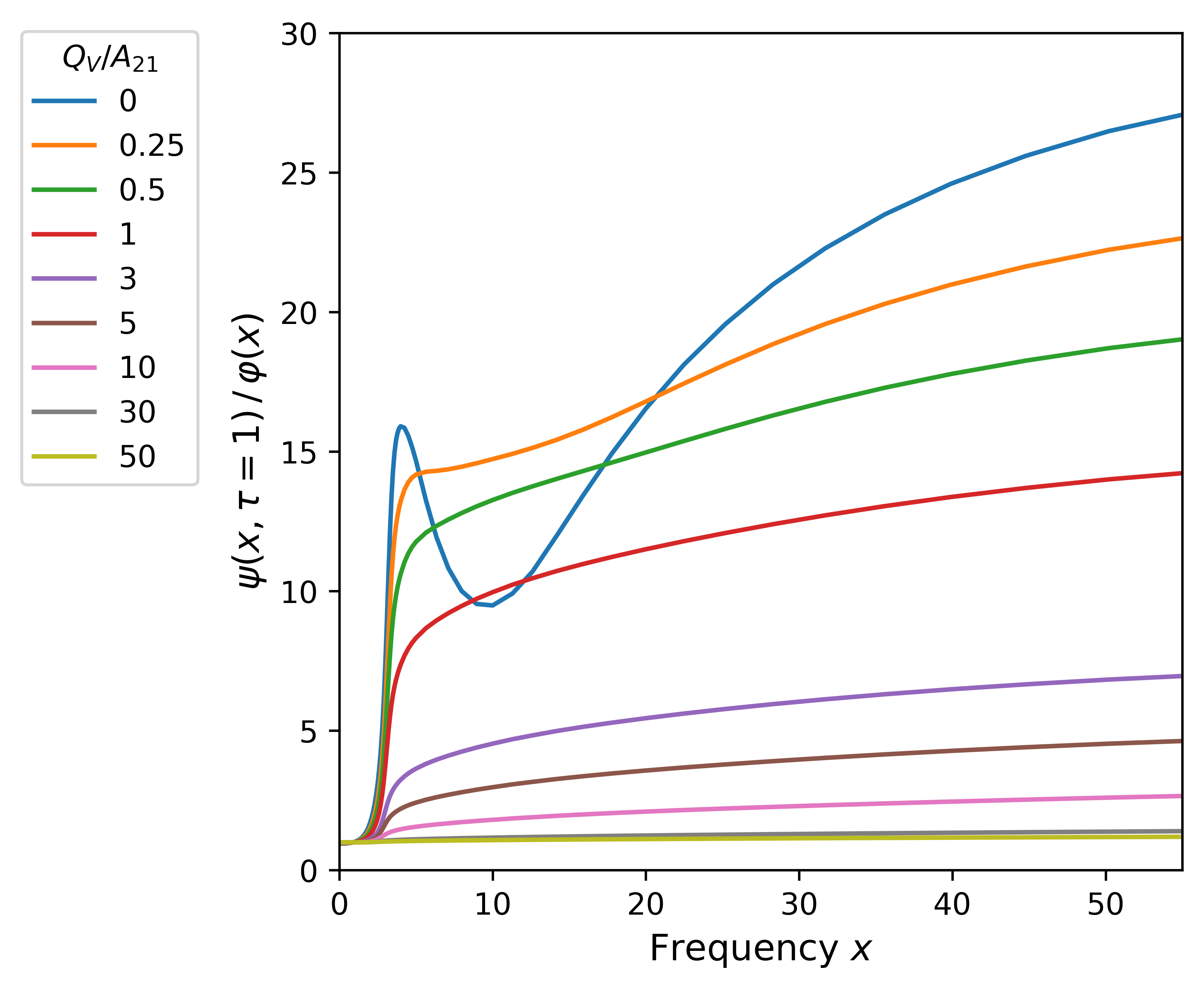}
  \caption{Dependence of the ratio of emission to absorption profile
  at $\tau=1$ (namely, $\psi(x,\tau=1)/\varphi(x)$) 
	on velocity-changing collisions. As expected, the emission
	profile approaches the CFR limit, namely
	$\psi(x,\tau)\to\varphi(x)$ with increasing values 
	of $Q_V/A_{21}$. }
  \label{fig6-psi-zeta}
\end{figure}

\begin{figure}[]
  \includegraphics[width=9cm, angle=0]{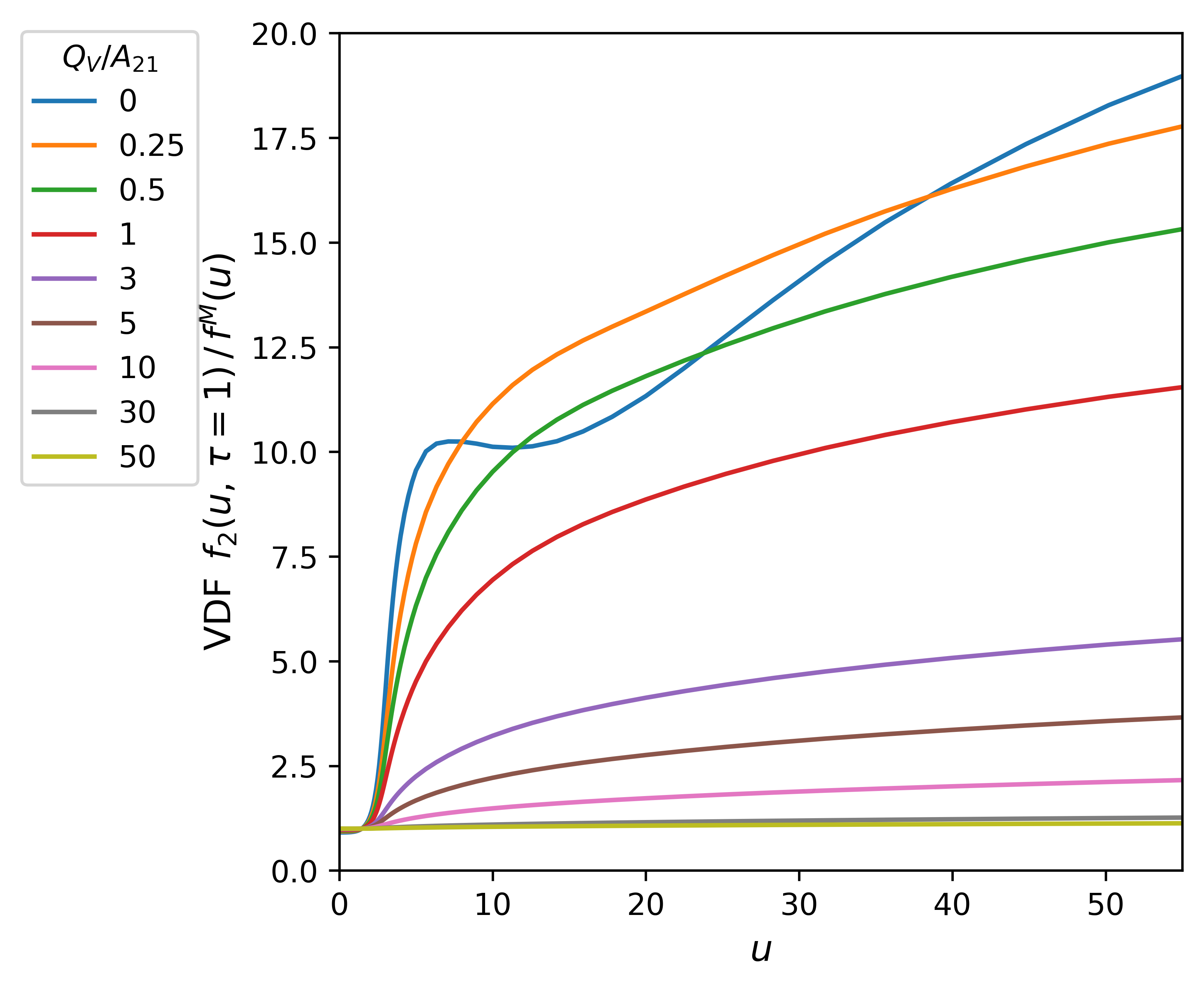}
  \caption{Dependence of the ratio of the VDF of the upper-level to Maxwellian 
  distribution at $\tau=1$ (namely, $f_2(u,\tau=1)/f^M(u)$) 
	on velocity-changing collisions. 
	With increasing values of $Q_V/A_{21}$ 
	the departure of the $f_2$ from Maxwellian initially increases 
	for $Q_V/A_{21}=0.25$ in the regime of intermediate 
	velocities and then decreases.}
  \label{fig7-f2-zeta}
\end{figure}

Unlike the standard non-LTE PFR formalism, the full non-LTE
formalism of \citet{ox86} takes into account the influence of
velocity-changing collisions characterized here by
$Q_V/A_{21}$. To bring out the impact of velocity-changing collisions, 
here we consider the extreme limit of $\alpha=0$, corresponding to the case of 
strong collisions \citep[in the kinetic sense; see][]{hc86}. When $\alpha=0$, 
the total elastic collision rate $Q_E$ gets its contribution solely from $Q_V$, 
which leads to simultaneous phase and velocity changes. In this respect, $Q_V$ 
here actually represents the effective velocity-changing collision rate. 
Figures~\ref{fig5-s-zeta}, \ref{fig6-psi-zeta}, and \ref{fig7-f2-zeta}, 
respectively, display the influence of $Q_V/A_{21}$ 
on the source function, emission profile, and the VDF of the upper level
at optical depth $\tau=1$. The model parameters used are the same 
as those for Figures~\ref{fig1-hum69-validate}--\ref{fig3-f2-hum69}, but we 
now vary $Q_V/A_{21}$ from $0$ to $50$ in much the same way as $\zeta$ was 
varied in Section~7 of \citet{psp23} for the CS case. Since we have chosen 
$\epsilon=10^{-4}$, the ratio $Q_I/A_{21}$ is relatively small, such that 
$\zeta$ (see Eq.~(\ref{zeta})) is nearly the same as $Q_V/A_{21}$. Thus 
our Fig.~\ref{fig5-s-zeta} is equivalent to Fig.~4 of \citet{psp23}, but for 
the case of broadened upper level. However, unlike the CS 
case, the frequency grid is much more extended in the present case. This is to 
take into account the fact that the absorption profile is now a Voigt 
function with rather broad damping wings. Like in the CS case, the source
function at $\tau=1$ approaches the CFR limit with increasing values
of $Q_V/A_{21}$ or $\zeta$ (see Fig.~\ref{fig5-s-zeta}). This is also 
in general the trend exhibited by the emission profile (see 
Fig.~\ref{fig6-psi-zeta}) and the VDF of the upper level (see 
Fig.~\ref{fig7-f2-zeta}). As discussed in Section~\ref{sec-coll-clarify}, the 
velocity-changing collisions are non-negligible in the lower solar atmosphere, 
wherein $Q_V/A_{21}$ or $\zeta$ may take moderate values when full 
non-LTE formalism has to be adopted for an accurate determination of the source 
function (cf. Fig.~\ref{fig5-s-zeta}) and the radiation field.

\subsection{Impact of phase-changing elastic collisions ($q_E/A_{21}$)}
\label{sec-ecoll}
\begin{figure}[]
  \includegraphics[width=9cm, angle=0]{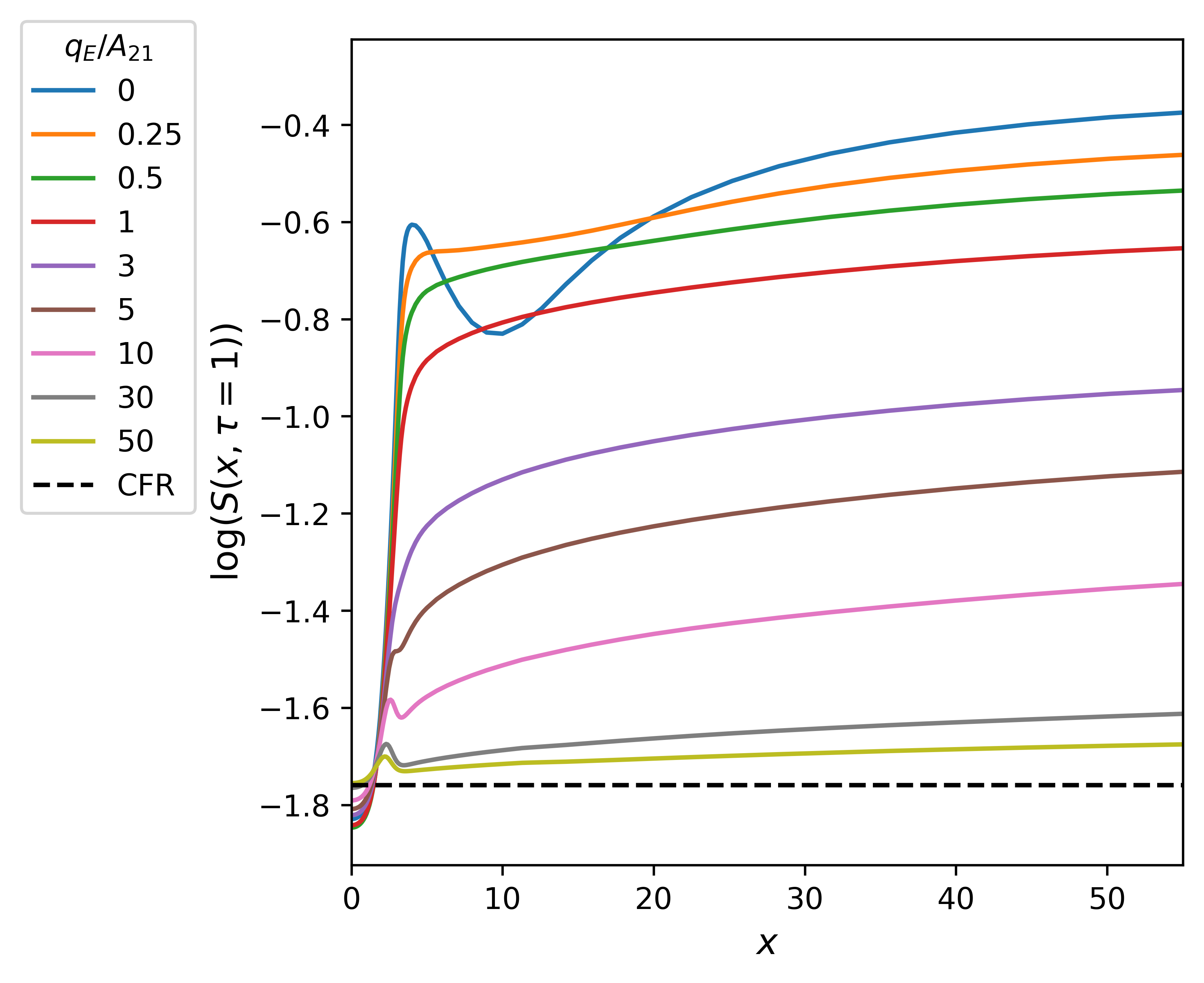}
	\caption{Influence of phase-changing elastic collision rate 
	$q_E/A_{21}$ on the normalized
        source function at $\tau=1$. As expected, the source function
	approaches the CFR limit (shown as horizontal dashed line) with
	increasing values of $q_E/A_{21}$.}
  \label{fig8-s-gebyr}
\end{figure}

\begin{figure}[]
  \includegraphics[width=9cm, angle=0]{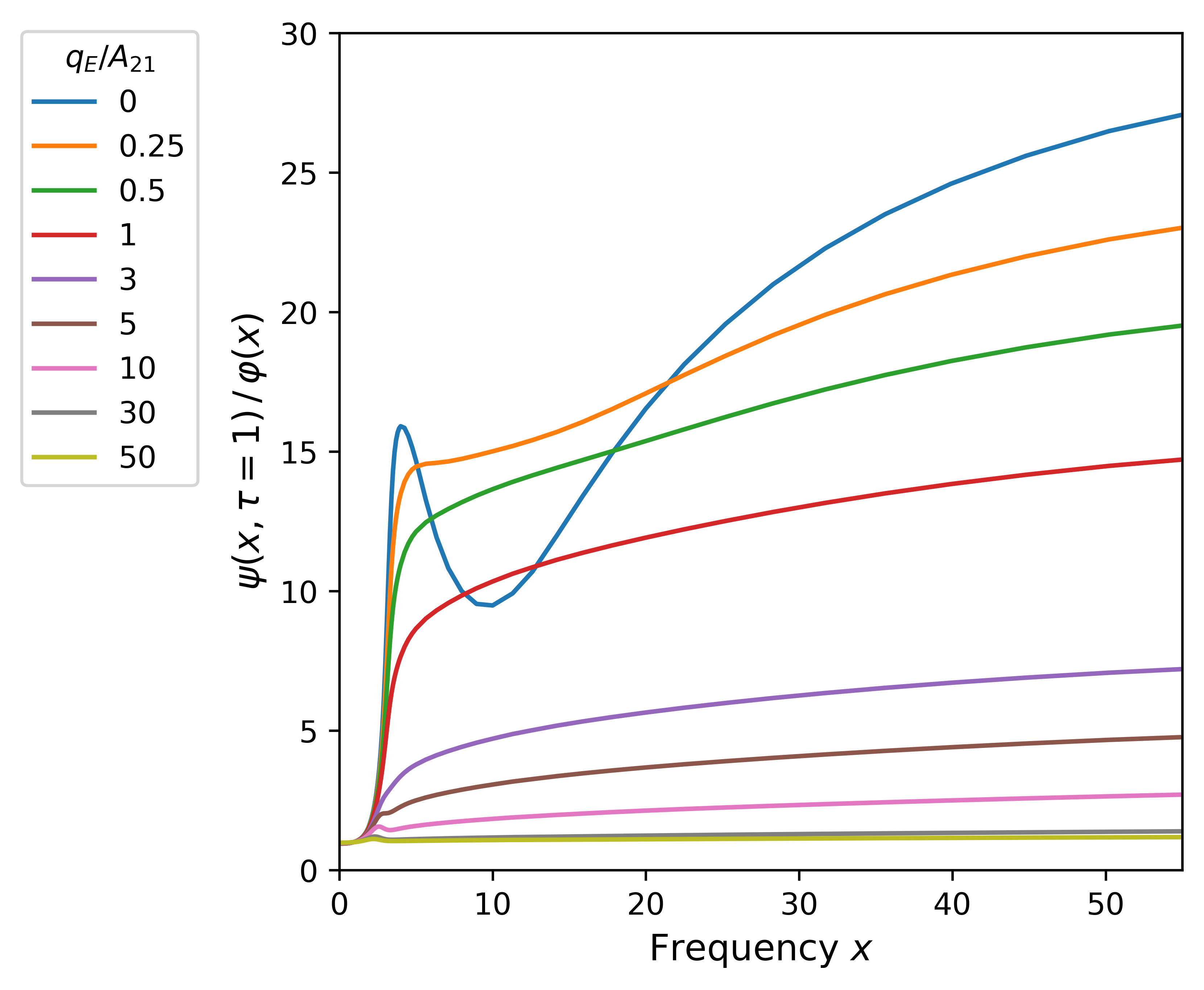}
  \caption{Dependence of the ratio of emission to absorption profile at
	$\tau=1$ (namely, $\psi(x,\tau=1)/\varphi(x)$) on varying 
	values of phase-changing elastic collision rate $q_E/A_{21}$ 
	(indicated in the figure legend). As expected, the emission profile 
	approaches the CFR limit (namely, $\psi(x,\tau)\to\varphi(x)$) 
	with increasing phase-changing elastic collision rate. }
  \label{fig9-psi-gebyr}
\end{figure}

\begin{figure}[]
  \includegraphics[width=9cm, angle=0]{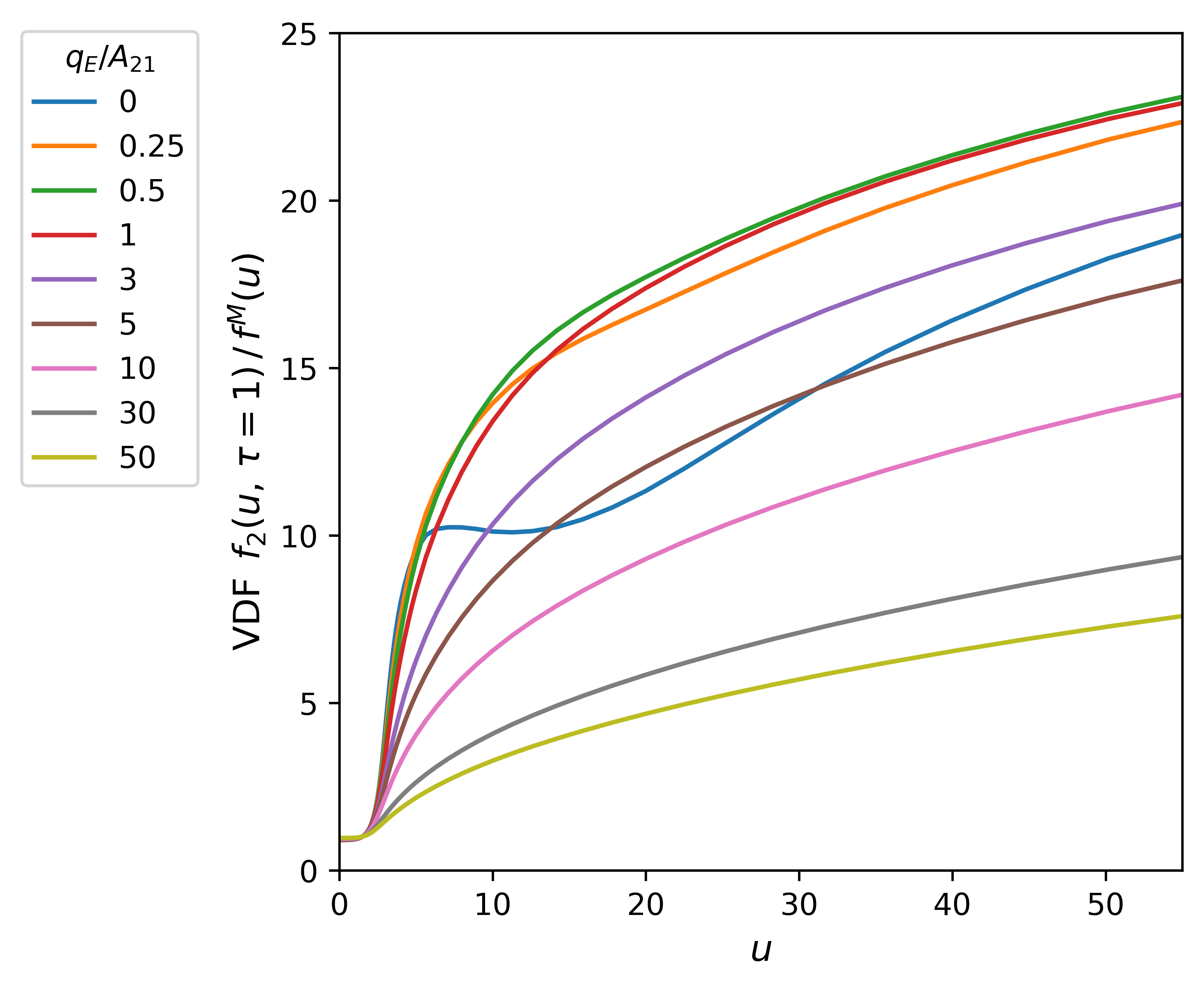}
  \caption{Dependence of the ratio of the VDF of the upper-level to Maxwellian 
  distribution at $\tau=1$ (namely, $f_2(u,\tau=1)/f^M(u)$) on 
	varying values of phase-changing elastic collision rate 
	$q_E/A_{21}$ (indicated in the figure legend). With increasing values 
	of $q_E/A_{21}$ the 
	the departure of the $f_2$ from Maxwellian initially increases 
	(until $q_E/A_{21}=0.5$) and then decreases.}
  \label{fig10-f2-gebyr}
\end{figure}

The phase-changing elastic collisions that are normally accounted for in 
spectral line formation theory through their effect on broadening the spectral 
line and leading to CFR in the atomic frame are characterized by 
$q_E/A_{21}$. By considering $\alpha=1$, here we present its impact on the 
source function, emission profile, and the VDF of the upper level at $\tau=1$. 
When $\alpha=1$, the total elastic collision rate $Q_E$ gets its contribution 
solely from the phase-changing collisions, which are basically weak collisions 
\citep{hc86}. The dependence of the source function 
on $q_E/A_{21}$ is known from the standard non-LTE PFR formalism. 
Basically with an increase in $q_E/A_{21}$ the source function 
approaches the CFR limit, which is indeed the case as seen from 
Fig.~\ref{fig8-s-gebyr}. This is true also for the emission profile, namely 
$\psi(x,\tau) \to \varphi(x)$ with increasing values of the 
phase-changing elastic collision rate (see Fig.~\ref{fig9-psi-gebyr}). 
As for the VDF of the upper level, its departure from the Maxwellian 
distribution initially increases until $q_E/A_{21}=0.5$ and 
then decreases with further increase in $q_E/A_{21}$ (see 
Fig.~\ref{fig10-f2-gebyr}). 

\section{Conclusions} 
\label{sec-conclu}
Full non-LTE radiative transfer, although formulated way back in
  the 1980's \citep{ox86}, remained largely unexplored because of the
complexity involved in its numerical implementation \citep[see
  however][who considered the limiting case of pure Doppler
  profile]{bos86,bos87,olgaetal87}. More recently, \citet{pp21} have
reconsidered this problem, and expressed its basic elements in
terms of the prevailing standard notations in this field of
research. \citet{psp23} then made a numerical implementation of this
formalism for the case of coherent scattering in the atomic frame
using the usual numerical iterative methods that are in use for the
standard non-LTE transfer problem. In the present paper, we have
  solved, for the first time, a full non-LTE radiative transfer
  problem considering the case of a two-level atom with infinitely
  sharp lower level and broadened upper level. 
For this purpose, we have applied, after suitable modifications, the
well-known operator perturbation methods developed for standard PFR 
models \citep[][see also \citealt{sj10,lambetal16}]{pa95}. 
We validate our iterative method against the standard non-LTE transfer problem
with angle-averaged $R_{II-A}$ PFR function \citep{hum62,hum69}. 
We illustrate the new quantities, namely, the emission profile and the
VDF of the upper level, and also make a comparison with the case of a
two-level atom with infinitely sharp lower and upper levels
\citep[namely, the coherent scattering case considered in][]{psp23}. 
We clearly demonstrate the influence of phase-changing elastic 
collisions ($q_E$, which lead to spectral line broadening and CFR 
in the atomic frame) and the velocity-changing collisions ($Q_V$) 
on the source function, emission profile, and the VDF of the upper level. 
In particular, we show that for moderate values of $Q_V/A_{21}$ (or 
equivalently $\zeta$; see Eq.~(\ref{zeta})), one has to adopt the full non-LTE 
formalism presented here to accurately determine the source function (see 
Fig.~\ref{fig5-s-zeta}) and the radiation field. Results presented in
  this paper would serve as benchmarks for future works in this topic
  (and would be made available upon request to corresponding author).

In the present paper, we show that in the absence of 
velocity-changing collisions, the full non-LTE formalism is equivalent to 
the standard non-LTE PFR formalism, thereby validating the use of numerically 
relatively simpler standard non-LTE PFR formalism. However, unlike the 
standard non-LTE formalism, the full non-LTE formalism can also account for 
the velocity-changing collisions, which may become significant in the lower 
solar atmosphere (see Section~\ref{sec-coll-clarify}). Given this, an accurate 
determination of velocity-changing collision rates for astrophysical 
applications become crucial. Until such calculations become available, the 
hard-sphere collision model would provide an excellent way to determine the 
cross-section for velocity-changing collisions.

  For computational simplicity, in the present paper we have 
  considered the angle-averaged emission profile (cf. Eq.~(\ref{psix})), 
  and thereby the angle-averaged redistribution functions (cf. 
  Eq.~(\ref{r121})). One of the near future goals would be to relax this 
  assumption, which would allow us to explore the angular dependence 
  of the VDF of the excited level.

The next crucial step will be to consider the full non-LTE
  transfer problem for multi-level atoms. In particular we intend to
  take forward this work by considering a three-level atom which would
  involve dealing with three distributions, one for the photon and two
  more for the excited atoms. Then another important step would be to
  relax the usual assumption of Maxwellian velocity distribution for
  the free electrons. 

\begin{acknowledgements}
  M.S. acknowledges the support from the Science and Engineering
  Research Board (SERB), Department of Science and Technology,
  Government of India via a SERB-Women Excellence Award research grant
  WEA/2020/000012. 
  We acknowledge the use of the high-performance computing facility 
 (\url{https://www.iiap.res.in/?q=facilities/computing/nova}) at the 
	Indian Institute of Astrophysics. M.S. would like to thank 
	Prof. Helene Frisch of OCA, Nice, France for useful discussions. 
	Authors thank an anonymous referee for constructive comments. 
	Authors are also thankful to the referee Prof. Ivan Hubeny 
	for carefully reviewing the work presented in this paper and for his 
	valuable comments.
\end{acknowledgements}

\bibliographystyle{aa}

\end{document}